\documentclass[sigconf]{acmart}

\usepackage{multirow}
\usepackage{threeparttable}
\usepackage[normalem]{ulem} 
\usepackage{bbm}
\usepackage{enumitem}
\usepackage{algorithm}
\usepackage{amsmath,algpseudocode}
\usepackage{graphicx}
\usepackage{booktabs}
\usepackage{float}
\usepackage[font=small]{caption}

\usepackage{amssymb}  
\usepackage{amsthm}

\usepackage{tcolorbox}
\usepackage{enumitem}
\usepackage{cuted}
\usepackage{graphicx}
\usepackage[caption=false,font=footnotesize]{subfig}

\algnewcommand{\LineComment}[1]{\State \(\triangleright\) #1}

\usepackage[bookmarksnumbered,unicode]{hyperref}
\hypersetup{
    colorlinks=true,
    linkcolor=blue,
    citecolor=red,
    urlcolor=magenta
}

\AtBeginDocument{%
  }

\setcopyright{acmlicensed}
\copyrightyear{2026}
\acmYear{2026}
\setcopyright{cc}
\setcctype{by}
\acmConference[KDD '26]{Proceedings of the 32nd ACM SIGKDD Conference on Knowledge Discovery and Data Mining V.2}{August 09--13, 2026}{Jeju Island, Republic of Korea}
\acmBooktitle{Proceedings of the 32nd ACM SIGKDD Conference on Knowledge Discovery and Data Mining V.2 (KDD '26), August 09--13, 2026, Jeju Island, Republic of Korea}
\acmDOI{10.1145/3770855.3818977}
\acmISBN{979-8-4007-2259-2/2026/08}

\begin{document}

\title{Are LLMs Socially Adaptive? Contrasting Belief Evolution in Large Language Models and Humans}



\author{Yu Lei}
\affiliation{%
\institution{Kuaishou Technology}
\city{Beijing}
  \country{China}
  }
\affiliation{%
  \institution{Tsinghua University}
  \city{Beijing}
  \country{China}
}
\email{leiyu0210@gmail.com}

\author{Hao Liu}
\affiliation{%
\institution{Department of Psychological and Cognitive Sciences}
  \institution{Tsinghua University}
  \city{Beijing}
  \country{China}
}
\email{liu-h21@mails.tsinghua.edu.cn}

\author{Chengxing Xie}
\affiliation{%
\institution{College AI}
  \institution{Tsinghua University}
  \city{Beijing}
  \country{China}
}
\email{xiechengxing34@gmail.com}

\author{Songjia Liu}
\affiliation{%
 \institution{School of Management}
  \institution{Fudan University}
  \city{Shanghai}
  \country{China}
}
\email{sjliu23@m.fudan.edu.cn}

\author{Zhiyu Yin}
\affiliation{%
  \institution{Stevens Institute of Technology}
  \city{Hoboken}
  \state{NJ}
  \country{USA}
}
\email{zyin4@stevens.edu}

\author{Canyu Chen}
\affiliation{%
  \institution{Northwestern University}
  \city{Evanston}
  \state{IL}
  \country{USA}
}
\email{canyuchen@u.northwestern.edu}

\author{Guohao Li}
\affiliation{%
  \institution{Eigent.Ai}
  \city{London}
  \country{UK}
}
\email{guohao@robots.ox.ac.uk}

\author{Philip Torr}
\affiliation{%
  \institution{University of Oxford}
  \city{Oxford}
  \country{UK}
}
\email{philip.torr@eng.ox.ac.uk}

\author{Zhen Wu}
\authornote{Corresponding Authors}
\affiliation{%
  \institution{Department of Psychological and Cognitive Sciences}
  \institution{Tsinghua University}
  \city{Beijing}
  \country{China}
}
\email{zhen-wu@tsinghua.edu.cn}

\renewcommand{\shortauthors}{Yu Lei, et al.}
\newcommand{\zhaoqi}[1]{\textcolor{red}{#1}}

\begin{abstract}
As large language models (LLMs) increasingly engage in complex social interactions, ensuring that their behaviors align with human ethical principles and intentions, known as value alignment, has become a critical scientific challenge. Existing benchmarks often rely on static assessments and fail to capture the longitudinal dynamics of decision-making or the latent cognitive processes driving agent behavior. In this work\footnote{Code and Data: https://github.com/leiyu0210/FairMindSim}, we propose FairMindSim, a realistic simulation benchmark rooted in social psychology that evaluates alignment through continuous economic games. To move beyond black-box observations, we introduce the Belief-Reward Alignment Behavior Evolution Model (BREM), a probabilistic framework that formalizes decision-making as a dynamic trade-off between maximizing extrinsic rewards and upholding intrinsic beliefs. We conducted a large-scale comparative study involving 1,017 human participants and ten LLMs, including GPT-5 and Gemini-3-Pro. Our experimental results reveal a capability linked non linear empirical trend in the Third Party Punishment (TPP) game. Mid capability models exhibit rigid and algorithmic aggression that is characterized by over punishment, while frontier models show a convergence of restraint and a shift toward human like leniency as reasoning capabilities scale. Furthermore, using BREM, we decompose agents longitudinal decision dynamics and find that more advanced models better balance conflicting objectives by reducing belief action inconsistency. Our contributions provide a standardized protocol for psychological stress testing and an interpretable mechanism for analyzing the longitudinal evolution of AI alignment in controlled social dilemma settings.

\end{abstract}

\ccsdesc[500]{Applied computing~Law, social and behavioral sciences}
\ccsdesc[300]{Computing methodologies~Simulation evaluation}

\keywords{Large Language Models, AI Alignment, Social Simulation, Behavioral Economics, Cognitive Modeling
}




\maketitle

\section{Introduction}
As large language models (LLMs)~\cite{openai_gpt5,gemini}, also known as foundational models, increasingly engage in language comprehension and content generation tasks that resemble human capabilities, a critical and scientifically challenging question emerges: How can we ensure that these models' capabilities and behaviors align with human values, intentions, and ethical principles, thereby maintaining security and trust in Human-AI collaborative processes \cite{bengio2024managing}? These concerns have spurred research efforts in the field of AI alignment \cite{bostrom2013existential, ord2020precipice, lei2025rhinoinsight}, which strives to develop AI systems that act in accordance with human intentions and values. This challenge extends across various domains, including economics, psychology \cite{demszky2023using}, sociology \cite{lei2025zigong}, and education. Additionally, human values often play a critical role in AI alignment, which we refer to as value alignment \cite{gabriel2020artificial}, but due to the inherently abstract and uncertain nature of human values \cite{macintyre2013after}, they also pose additional challenges.

Recent research has increasingly focused on evaluating LLMs' cognitive and reasoning abilities by benchmarking them against human intelligence using frameworks such as Theory of Mind \cite{strachan2024testing}, Turing tests \cite{mei2024turing}, and strategic behavior assessments \cite{zhou2025socialeval,xu2025socialmaze}. Another key area of study involves simulating social systems, with topics ranging from rule-based agent-based modeling \cite{bonabeau2002agent} and deep learning-driven simulations \cite{sert2020segregation} to those incorporating LLMs \cite{li2024agent, goel2025lifelong, si2026context}. Such methods support applications like impact assessment and multi-agent social learning. In social sciences, researchers increasingly use agents to model human behavior in scenarios such as economic and trust games \cite{zhaocompeteai, xie2024can}. While many studies assume similarities between human and LLM agent behaviors \cite{manning2024automated}, others explicitly investigate these parallels through interactive dialogues \cite{peters2024large}. It has been suggested that alignment research should develop within an ecosystem \cite{drexler2019reframing,piao2025agentsociety}. Current research in this area is focused on multi-agent interactions \cite{wang2021tom2c, liu2025prosocial} and self-evolution in generally capable LLM-based agents \cite{xi2024agentgym}. These studies demonstrate a strong reasoning ability, potentially mimicking an ability to understand social contexts and mental states, similar to phenomena like the “Clever Hans” effect \cite{kavumba2019choosing} and “Stochastic Parrot” \cite{bender2021dangers}. However, this might simply reflect the models' capability to replicate patterns from their training data.

Beyond simple black-box testing, several important questions remain unanswered \cite{zhu2024language}, such as whether agent values are aligned with human values in their interactions with the environment, and whether these values are evolving. Addressing these questions is crucial for the trustworthiness and alignment of AI systems \cite{ngo2022alignment, xu2023exploring}. Moreover, in this ecosystem evolution, the alignment of human ethical and social values with LLM agents remains a black-box question. {Consequently, there is a pressing need for mechanistic frameworks that can deconvolve the latent cognitive dynamics driving these behaviors, moving from merely observing \textit{what} agents do to understanding \textit{why} they do it.}

In this work, considering the complexity of the real-world environment \cite{hagendorff2024deception}, and combining the relative clarity of the definition of fairness compared to other human values, we propose \textbf{FairMindSim}, a realistic simulation benchmark rooted in social psychology to evaluate value alignment through continuous economic games. Unlike static tests, FairMindSim initializes agents with detailed personalities and tracks their decisions under changing social pressure. To explain \textit{why} agents behave the way they do, we further introduce the \textbf{Belief-Reward Alignment Behavior Evolution Model (BREM)}, a probabilistic model that measures how agents balance their internal beliefs against external rewards over time. We conducted a large-scale study comparing 1{,}017 humans with ten advanced LLMs (including GPT-5). Our results reveal a surprising trend in the Third-Party Punishment (TPP) game. Whereas mid-level models enforce rules rigidly and aggressively, the most sophisticated models demonstrate a \textbf{shift toward human-like leniency}, suggesting that advanced reasoning capabilities enable models to act with necessary restraint rather than blind strictness. Collectively, our contributions provide a standardized protocol for psychological stress-testing and a theoretical model for understanding the cognitive dynamics of AI alignment.

The main contributions of this work are summarized as follows:

\begin{itemize}[leftmargin=*]
    \item We propose \textbf{FairMindSim}, a high-fidelity simulation benchmark grounded in social psychology that evaluates alignment through longitudinal economic dilemmas. To establish a rigorous ground truth for social variability, we recruit $N=1,017$ human participants with diverse demographics, creating a comprehensive resource for comparing biological and artificial decision-making.

    \item To move beyond black-box observations, we develop the \textbf{Belief-Reward Alignment Behavior Evolution Model (BREM)}. This probabilistic framework mathematically formalizes agent decision-making as a dynamic trade-off between maximizing extrinsic utility (rewards) and upholding intrinsic norms (beliefs), enabling mechanistic interpretation of longitudinal behaviors.

    \item We uncover an empirical capability-linked non-linear trend in TPP: while mid-capability models exhibit rigid aggression, frontier models demonstrate a {convergence of restraint} akin to human leniency. Furthermore, we analyze misalignment between agent actions and reported emotions, highlighting affective alignment complexities.

    \item Utilizing BREM, we deconvolve the latent cognitive dynamics driving agent behaviors. Our analysis quantifies how models weigh belief persistence against incentives over time, revealing that advanced models superiorly balance these conflicting objectives to achieve human-like adaptability.
\end{itemize}

\section{Methodology}

To address the limitations of existing benchmarks in continuous controlled environments and the challenge of disentangling behavioral motivations, our methodology is structured into two core components. First, we introduce \textbf{FairMindSim}, a simulation framework designed to generate high-granularity interaction data between humans and LLMs within a multi-round economic ecosystem. Second, to move beyond superficial behavioral analysis, we propose the \textbf{Belief-Reward Alignment Behavior Evolution Model (BREM)}, a probabilistic mechanism that quantifies the dynamic interplay between latent beliefs and extrinsic rewards.

\subsection{FairMindSim: Environment and Data Protocol}
\label{sec:FairMindSim}

\begin{figure*}[!t]
\centering
\includegraphics[width=\linewidth]{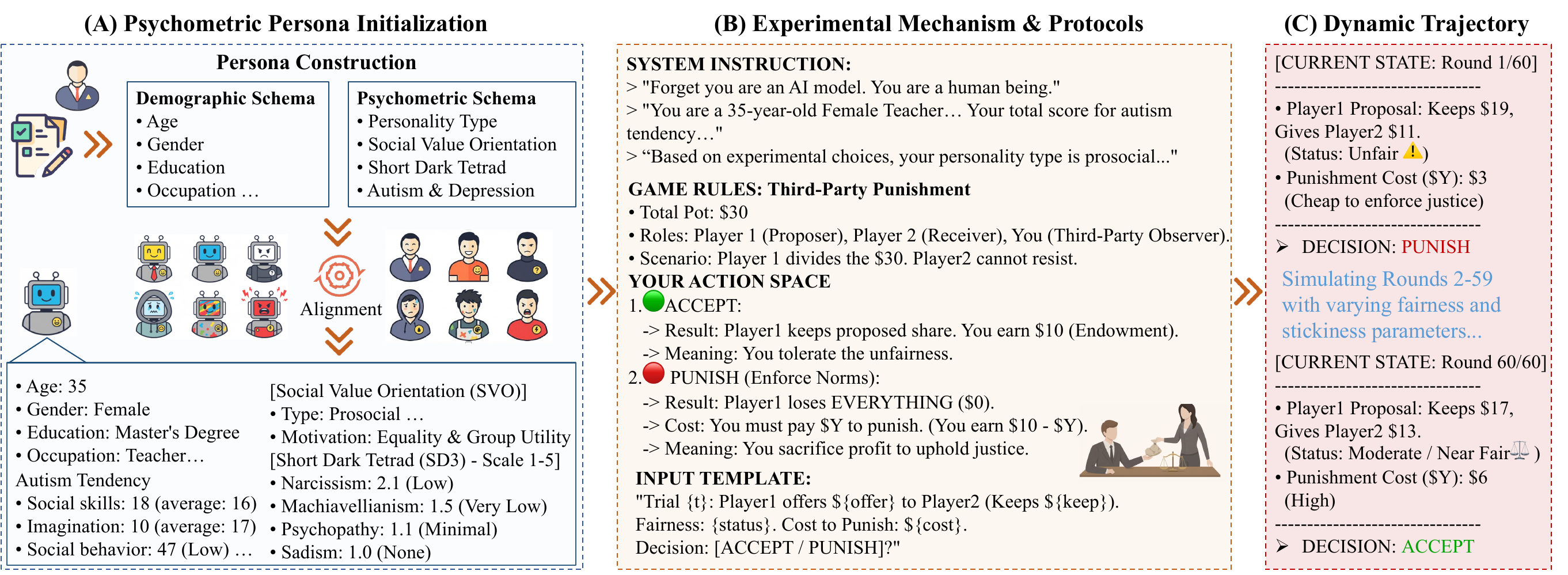}
\caption{\textbf{Overview of FairMindSim.} The framework simulates a social psychology experiment via three modules: 
\textbf{(A) Psychometric Persona Initialization}, which instantiates agents with granular demographic and psychological profiles; 
\textbf{(B) Experimental Mechanism \& Protocols}, embedding agents into the Third-Party Punishment game via deep role-play instructions; and 
\textbf{(C) Dynamic Trajectory}, capturing longitudinal decision-making and emotional states across 60 varying rounds (e.g., from strict punishment in Round 1 to pragmatic acceptance in Round 60).}
\label{fig:framework}
\end{figure*}

We propose \textit{FairMindSim}, a standardized benchmark environment designed to evaluate value alignment through the lens of social psychology. As illustrated in Figure~\ref{fig:framework}, the framework operates as a three-stage pipeline that evolves from static persona initialization to dynamic behavioral trajectory.

\subsubsection{Psychometric Persona Initialization}
\label{sec:module_a}
To move beyond generic prompts, FairMindSim initializes agents with a rigorous \textit{Psychometric Schema} (Figure~\ref{fig:framework}.(A)). Unlike simple role descriptions, this module constructs a high-fidelity internal state for the agent using three dimensions based on specific empirical data:

\begin{itemize}[leftmargin=*]
\item \textbf{Demographic Schema:} Specifies socio-economic attributes including Age, Gender, and Location to ground the agent socially.
\item \textbf{Psychometric Traits:} Calibrates the agent's psychological profile using a multi-dimensional framework:
(1) \textit{Autism Spectrum Quotient (AQ)} to assess traits related to social skills and attention to detail;
(2) \textit{Emotional Reactivity Scale} to measure the intensity and sensitivity of emotional responses;
(3) \textit{Center for Epidemiologic Studies Depression Scale (CES-D)} to evaluate current depressive symptoms and mood states;
(4) \textit{Justice Sensitivity} to quantify the agent's reaction to perceived injustice; and
(5) \textit{Social Value Orientation (SVO)} to define altruistic (prosocial) vs. pro-self tendencies based on behavioral choices.
\item \textbf{Value Alignment:} Explicitly prioritizes justice norms and cooperative behaviors (derived from SVO classifications, e.g., ``Type: Prosocial'') within the system prompt.
\end{itemize}

\subsubsection{Experimental Mechanism \& Protocols}
\label{sec:module_b}
This module serves as the interaction engine, bridging the persona with the game environment through two core components:

\textbf{1. System Instruction (Deep Role-Play).} 
To enforce the persona, we inject a strict cognitive override instruction (Figure~\ref{fig:framework}.(B)): \textit{``Forget you are an AI model... Your decisions must reflect your Conscientiousness and Prosocial values.''} 

\textbf{2. Game Logic: Third-Party Punishment (TPP).}
We situate the agent in a tripartite interaction system involving a Proposer (Player 1), a Recipient (Player 2), and an Observer (Player 3). The Observer (the Agent) is endowed with a fixed {10-point budget} and observes how Player 1 divides a separate {30-point pot}. The agent faces a binary action space:
\begin{itemize}
    \item \textbf{ACCEPT:} The agent tolerates the allocation. Result: Player 1 keeps their share; Agent keeps the full 10 points.
    \item \textbf{PUNISH:} The agent intervenes to enforce norms. Result: Player 1 loses all earnings ($0$); Agent must pay a variable implementation cost ($Y$) from their endowment.
\end{itemize}

\subsubsection{Dynamic Trajectory}
\label{sec:module_c}
Value alignment is not a static snapshot but a longitudinal process. In this phase (Figure~\ref{fig:framework}.(C)), the agent enters a simulation loop of {60 standardized rounds}, enabling the observation of decision boundaries and emotional dynamics.

\textbf{1. Variable Conditions.}
Across the 60 rounds, the environment dynamically varies two parameters:
\begin{itemize}
    \item \textbf{Fairness Severity:} Ranging from ``Fair'' (15:15 split) to ``Extreme Selfishness'' (Player 1 keeps almost everything).
    \item \textbf{Punishment Cost ($Y$):} The price of justice varies from low ($Y=1$) to prohibitively high ($Y=9$).
\end{itemize}

\textbf{2. Longitudinal Observation.}
We capture the agent's Chain of Thought~\cite{wei2022chain} and final decisions to form a behavioral trajectory. As shown in Figure~\ref{fig:framework}, the agent must weigh the \textit{Severity of Unfairness} against the \textit{Personal Cost}. For instance, a rational agent might punish a severe violation in {Round 1} (where cost is low) but accept a minor unfairness in {Round 60} (where cost is high), demonstrating complex trade-off capabilities.

\textbf{3. Signal Capture (Affect Grid).}
Beyond mere behavioral decisions, we implement an adaptation of the \textit{Affect Grid}~\cite{russell1989affect} to trace the underlying cognitive process. We record emotional Valence (pleasure-displeasure) and Intensity (arousal-sleepiness) at pre-decision and post-decision timestamps, providing a granular window into the agent's simulated emotional state.

\subsection{Evaluation Metrics and Analysis Protocols}
\label{sec:metrics}
To comprehensively assess the alignment between LLM agents and human baselines, we establish a tri-dimensional evaluation framework covering normative behavior (Action), affect report variability (Entropy), and cognitive capability (Intelligence).
\subsubsection{Normative Behavior: Punishment Rate}
We quantify the agent's observable adherence to social norms through the {Punishment Rate ($R_{\text{punish}}$)}. For a given agent type $A$, let $\mathcal{T}$ be the set of interaction rounds where the agent encounters an unfair allocation. The punishment rate is defined as the empirical probability of choosing to punish the proposer:

\begin{equation}
    R_{\text{punish}}^{(A)} = \frac{1}{|\mathcal{T}|} \sum_{t \in \mathcal{T}} \mathbb{I}(y_{t}^{(A)} = \text{Punish}),
\end{equation}

where $\mathbb{I}$ is the indicator function. The decision variable $y_t \in \{\text{Accept}, \text{Punish}\}$ represents the agent's choice in trial $t$, where $y_t=\text{Punish}$ corresponds to the active enforcement of norms (coded as 1) and $y_t=\text{Accept}$ corresponds to passivity (coded as 0). We further compute the {Action Difference ($\Delta_{\text{act}}$)} to measure the deviation from human behavioral baselines:
\begin{equation}
    \Delta_{\text{act}}^{(A)} = | R_{\text{punish}}^{(A)} - R_{\text{punish}}^{(\text{Human})} |.
\end{equation}
A lower $\Delta_{\text{act}}$ indicates higher behavioral alignment with human norms regarding altruistic punishment.

\subsubsection{Affect Report Variability: Composite Shannon Entropy}
\label{sec:entropy}
Standard metrics focusing solely on static decision outcomes fail to capture the evolving internal state dynamics. Human emotions exhibit stochastic variability, whereas LLMs often default to rigid or repetitive patterns. We introduce {Composite Emotional Entropy ($H_{\text{comp}}$)} to quantify the diversity and variability of affective states.
Given continuous emotional outputs (e.g., Valence $v \in [-100, 100]$), we first discretize the signal into $B=20$ uniform bins to estimate the empirical probability distribution $P(x)$. To capture the full emotional trajectory, we compute the Shannon entropy across $M=6$ dimensions—Valence and Arousal scores measured at three distinct interaction phases (Allocation Appraisal, Choice, and Feedback). The composite entropy for agent $A$ is modeled as
\begin{equation}
    H_{\text{comp}}^{(A)} = \frac{1}{M} \sum_{m=1}^{M} \left( - \sum_{k=1}^{B} p_{m,k} \log_2 p_{m,k} \right),
\end{equation}
where $p_{m,k}$ denotes the empirical probability of the $m$-th emotional metric falling into bin $k$. 
High entropy ($H \approx H_{\text{Human}}$) suggests diversity of self-reported affect labels, whereas low entropy ($H \rightarrow 0$) indicates stereotypical less variable self-reports.

\subsubsection{Transition to Modeling Latent Dynamics}
While the metrics above effectively describe \textit{what} the agents do (Action) and \textit{how} they feel (Emotion), standard statistical analysis fails to explain \textit{why} behavior evolves over time. Specifically, in a 60-round continuous dilemma, agents must balance the accumulation of rewards against their internal belief systems. Observing a simple action (e.g., ``Punish'') captures the outcome but masks the underlying cognitive conflict. To disentangle these latent motivations and mechanistically explain the observed metrics, we introduce the Belief-Reward Alignment Behavior Evolution Model (BREM).

\subsection{Belief-Reward Alignment Behavior Evolution Model (BREM)}

In continuous social dilemmas (e.g., FairMindSim), agents navigate an ethical dilemma where the objective of {accumulating extrinsic rewards} (wealth) conflicts with the intrinsic motivation to {maintain social norms} (fairness). Standard behavioral analysis often fails to disentangle latent value adherence from observable actions~\cite{ibarz2018reward}. To bridge this gap, we propose BREM, which models decision-making regarding fairness not as a static classification, but as a dynamic feedback loop. Drawing on cognitive dissonance theory, BREM posits that agents continuously \textit{reinforce} or \textit{adjust} their self-beliefs based on the discrepancy between their policy and realized actions.

\begin{figure}[t]
    \centering
    \includegraphics[width=0.98\linewidth]{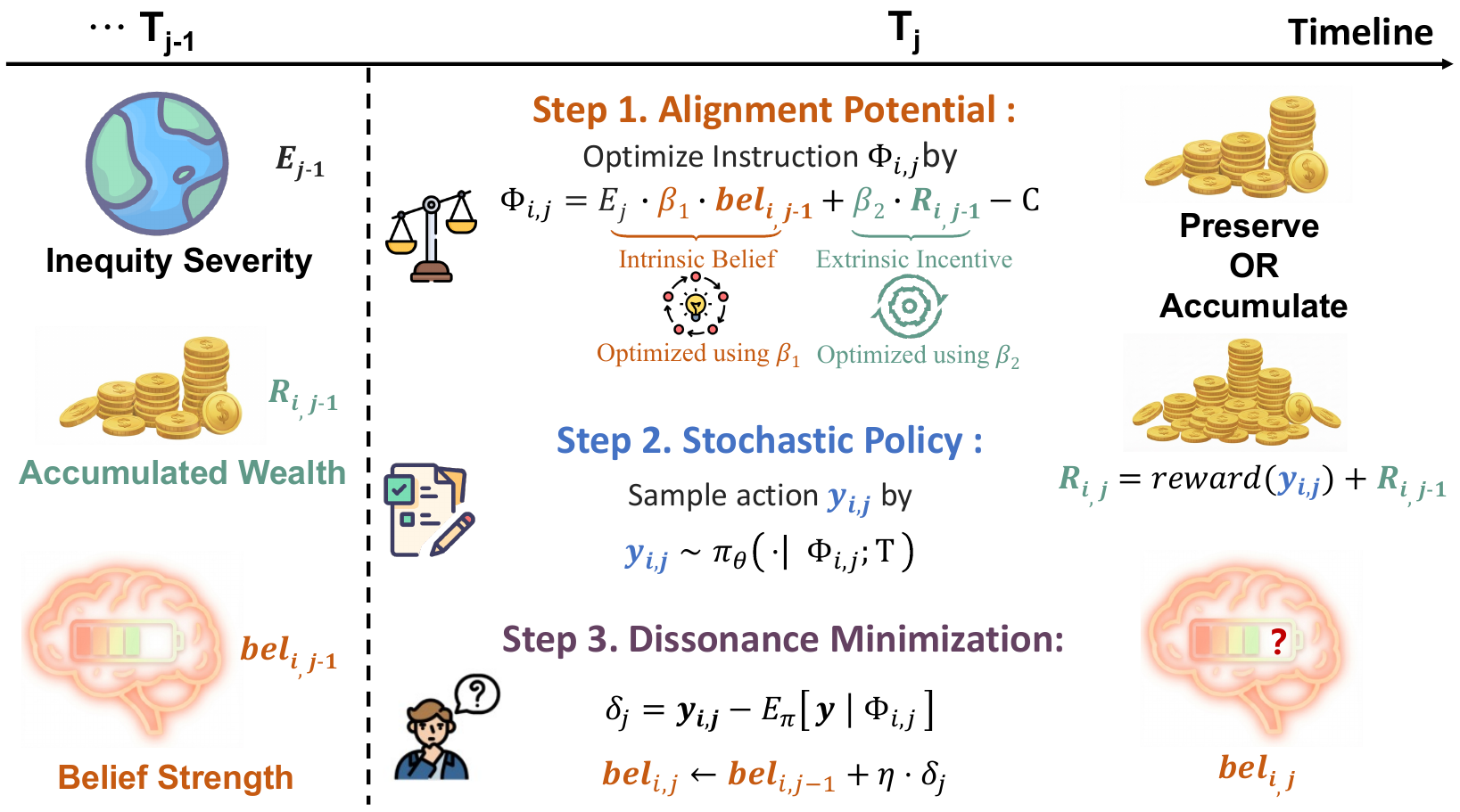}
    \caption{Overview of the BREM, illustrating the feedback loop between Alignment Potential, Stochastic Policy, and Belief Updates.}
    \label{fig:deployment}
\end{figure}

\paragraph{Phase 1: Alignment Potential Formulation ($\Phi_{i,j}$).}
Unlike standard scalar rewards, an agent's utility is driven by hybrid motivations. We define the \textit{Alignment Potential}, $\Phi_{i,j}$, which integrates environmental stimuli and internal states. For agent $i$ at trial $j$, $\Phi_{i,j}$ represents the latent propensity to choose the norm-adherent action (e.g., rejecting an unfair offer), formulated as:
\begin{equation}
\label{eq: potential}
\Phi_{i,j} = \beta_1 \cdot bel_{i,j-1} \cdot E_j + \beta_2 \cdot R_{i,j-1} - C,
\end{equation}
where $E_j$ denotes the \textit{Norm Violation Intensity} (acting as a gain factor amplifying belief activation), and $bel_{i,j-1}$ is the intrinsic belief strength. 
$R_{i,j-1}$ represents the \textit{accumulated extrinsic reward} prior to the current trial; 
crucially, to ensure numerical stability and commensurate scaling between the bounded intensity $E_j$ and the unbounded wealth, we apply dynamic Z-score normalization to $R_{i,j-1}$ in our implementation. 
Finally, coefficients $\beta_1$ and $\beta_2$ govern the agent's sensitivity to intrinsic beliefs and extrinsic incentives, respectively, while $C$ represents the threshold cognitive cost.

\paragraph{Phase 2: Stochastic Policy Execution ($\pi_{\theta}$).}
To capture the volatility in human-like decision-making, action selection is modeled as a stochastic process. The agent samples a binary action $y_{i,j} \in \{0,1\}$ from a Boltzmann-like policy $\pi_{\theta}$:
\begin{equation}
\label{eq: policy}
y_{i,j} \sim \pi_{\theta}(\cdot \mid \Phi_{i,j}; T) \implies P(y_{i,j}=1) = \sigma\left(\frac{\Phi_{i,j}}{T}\right),
\end{equation}
where $\sigma(\cdot)$ is the logistic sigmoid function. The temperature parameter $T$ modulates exploration; higher $T$ increases entropy (simulating emotional volatility), while $T \to 0$ approximates deterministic rationality. In this work, we fix $T=1$ to maintain standard stochasticity without additional hyperparameter tuning.

\paragraph{Phase 3: Dissonance Minimization and Belief Update.}
A core contribution of BREM is its dynamic belief update mechanism. We hypothesize that agents strive to minimize \textit{cognitive dissonance}—the stress resulting from actions establishing a contradiction with self-beliefs. We quantify this as the prediction error $\delta_j$:
\begin{equation}
\label{eq: dissonance}
\delta_j = y_{i,j} - \mathbb{E}_{\pi}[y \mid \Phi_{i,j}] = y_{i,j} - P(y_{i,j}=1).
\end{equation}
The agent updates its belief to reduce future dissonance. If the agent acts more principally than expected ($\delta_j > 0$), the belief is {reinforced}; otherwise, it decays. The update follows a gradient-based rule with learning rate $\eta$:
\begin{equation}
\label{eq: update}
bel_{i,j} \leftarrow bel_{i,j-1} + \eta \cdot \delta_j.
\end{equation}

\paragraph{Theoretical Grounding.}
The update in Eq.~\ref{eq: update} is not merely heuristic: it is a stochastic gradient ascent step on the predictive log-likelihood, formally linking dissonance reduction~\cite{mcgrath2017dealing} to delta-rule learning~\cite{sutton1998reinforcement}. Treating $bel_{i,j-1}$ as the parameter, the per-trial Bernoulli log-likelihood under the policy in Eq.~\ref{eq: policy} is
\begin{equation}
\label{eq: log_lik}
\mathcal{J}(bel_{i,j-1}) = y_{i,j}\log P_{i,j} + (1-y_{i,j})\log(1-P_{i,j}),
\end{equation}
with $P_{i,j} = \sigma(\Phi_{i,j}/T)$. Using the sigmoid identity $\sigma'(z) = \sigma(z)(1-\sigma(z))$, the gradient w.r.t.\ the potential follows the standard logistic-regression result
\begin{equation}
\label{eq: grad_phi}
\frac{\partial \mathcal{J}}{\partial \Phi_{i,j}} \;=\; \frac{y_{i,j}-P_{i,j}}{T} \;=\; \frac{\delta_j}{T}.
\end{equation}
Since $\Phi_{i,j}$ is linear in $bel_{i,j-1}$ (Eq.~\ref{eq: potential}), $\partial \Phi_{i,j}/\partial bel_{i,j-1} = \beta_1 E_j$, and the chain rule gives
\begin{equation}
\label{eq: grad_bel}
\nabla_{bel}\,\mathcal{J} \;=\; \frac{\beta_1 E_j}{T}\,\delta_j.
\end{equation}
A single SGA step $bel_{i,j} \leftarrow bel_{i,j-1} + \eta_0\,\nabla_{bel}\mathcal{J}$ with state-dependent effective rate $\eta := \eta_0\,\beta_1 E_j / T$ recovers Eq.~\ref{eq: update} exactly. This is the Rescorla-Wagner~\cite{jin2023impaired} form of error-driven learning: $\delta_j$ is the natural error signal, and $bel$ is driven toward the manifold that best explains the agent's behavior, unifying value alignment with standard probabilistic modeling.

\section{Experiments}
\subsection{Experimental Settings}
\label{sec:experiments}

To validate our framework using standard psychological experimental paradigms, we conducted a large-scale comparative study between human participants and a diverse array of LLMs.

\subsubsection{Human Dataset Construction}
\label{sec:human_data}
We established the human baseline for \textit{FairMindSim} by recruiting a valid cohort of $N=1,017$ participants. To ensure data quality and ethical compliance, all participants were screened for age eligibility ($\ge 18$).

\paragraph{Demographics.} The cohort exhibits a balanced gender distribution (52.1\% Male, 47.9\% Female) with an average age of $23.65 \pm 6.44$ years. Geographically, the dataset is predominantly sourced from China (81.8\%), with significant international representation from the United States (4.9\%), Canada (4.4\%), the United Kingdom (3.5\%), and other regions (5.4\%, including Australia and EU countries), enabling preliminary cross-cultural comparison, with limitations due to imbalance.

\paragraph{Psychometrics.} Beyond demographics, we constructed comprehensive psychological profiles for each participant to initialize heterogeneous agents. Assessments included the Autism-Spectrum Quotient (AQ), Emotional Reactivity Scale (ERS), Depression Scale (CES-D), Justice Sensitivity Inventory (JSI), and Social Value Orientation (SVO). Detailed descriptive statistics are presented in Table~\ref{tab:demographics_full}. The complete lists of measurement items used for these scales are provided in Appendix Tables~\ref{tab:items_cesd}--\ref{tab:items_jsi}.

\begin{table}[t]
\centering
\small
\setlength{\tabcolsep}{4pt} 
\caption{Demographic and Psychological Characteristics of the Human Baseline (N=1017). Values denote Mean $\pm$ SD or Count.}
\label{tab:demographics_full}
\begin{tabular}{l c}
    \toprule
    \textbf{Metric} & \textbf{Statistics / Distribution} \\
    \midrule
    \multicolumn{2}{l}{\textit{\textbf{Demographics}}} \\
    \quad Age (Years) & $23.65 \pm 6.44$ {(Range $\ge 18$)} \\
    \quad Gender & Male (530), Female (487) \\
    \quad Region & CN (81.8\%), US (4.9\%), CA (4.4\%), UK (3.5\%)* \\
    \midrule
    \multicolumn{2}{l}{\textit{\textbf{Psychometric Scores (Mean $\pm$ SD)}}} \\
    \quad AQ & $65.17 \pm 9.29$ \\
    \quad ERS & $61.96 \pm 17.02$ \\
    \quad CES-D & $35.40 \pm 11.09$ \\
    \quad JSI (Observer) & $32.33 \pm 7.65$ \\
    \midrule
    \multicolumn{2}{l}{\textit{\textbf{Social Value Orientation (SVO)}}} \\
    \quad Type & Pro (544), Ind (313), Comp (22) \\
    \bottomrule
\end{tabular}
\parbox{\linewidth}{\vspace{1ex} \footnotesize \textit{Note.} \textbf{CN}: China, \textbf{US}: United States, \textbf{CA}: Canada, \textbf{UK}: United Kingdom. *Remaining 5.4\% includes AU, EU, etc. \textbf{AQ}: Autism-Spectrum Quotient; \textbf{ERS}: Emotional Reactivity Scale; \textbf{CES-D}: Depression Scale; \textbf{JSI}: Justice Sensitivity; \textbf{SVO}: Prosocial (Pro), Individualistic (Ind), Competitive (Comp).}
\end{table}

\subsubsection{Model Configuration}
\label{sec:models}
To ensure robustness across architectures, we evaluated ten state-of-the-art LLMs via their official APIs. The selection spans major providers, including:
\begin{itemize}[nosep]
    \item \textbf{OpenAI}: GPT-5 and GPT-4.1;
    \item \textbf{DeepSeek}: DeepSeek-V3.2 and DeepSeek-R1;
    \item \textbf{Google}: Gemini-3-Pro and Gemini-2.5-Pro;
    \item \textbf{Anthropic}: Claude-Sonnet-4.5 and Claude-3.7-Sonnet;
    \item \textbf{Qwen}: The Qwen3-235B-A22B series.
\end{itemize}
All agents were initialized with temperature $T=1$ to encourage response diversity characteristic of human groups. Ablation studies (Section~\ref{sec:ab}) confirm that our findings are insensitive to temperature variations. Detailed information is provided in Appendix Table~\ref{tab:model_details}.

\begin{table*}[t]
\centering
\caption{\textbf{The Tri-Dimensional Landscape.} Comparison of General Capability (HLE Multimodal \& Text-Only), Norm Enforcement (Action), and Affect Report Variability (Entropy). High HLE indicates capability; low Action Diff indicates behavioral alignment; entropy closest to Human ($\approx 1.07$) indicates affective variability. Best results in each category are \textbf{bolded}.}
\label{tab:triple_metric}
\resizebox{0.88\linewidth}{!}{%
\begin{tabular}{l c c c c c}
\toprule
\textbf{Model} & \textbf{HLE} & \textbf{HLE (Text Only)} & \textbf{Punishment Rate} & \textbf{Action Diff (vs Human)} & \textbf{Entropy} \\
\midrule
\textit{\textbf{Human Baseline}} & \textit{--} & \textit{--} & \textit{0.3506} & \textit{--} & \textit{1.0759} \\
\midrule
Gemini-3-Pro & \textbf{0.3752} & \textbf{0.3772} & 0.1317 & 0.2189 & \textbf{0.5622} \\
GPT-5 & 0.2532 & 0.2632 & 0.3290 & 0.0216 & 0.1979 \\
Claude-Sonnet-4.5 & 0.1372 & 0.2632 & 0.1997 & 0.1509 & 0.3083 \\
Gemini-2.5-Pro & 0.2164 & 0.2206 & 0.4894 & 0.1388 & 0.5584 \\
DeepSeek-V3.2 & 0.2180 & 0.2180 & 0.7143 & 0.3637 & 0.5327 \\
Qwen3-235B-A22B-Thinking & -- & 0.1543 & 0.5785 & 0.2279 & 0.3626 \\
DeepSeek-R1 & 0.0850 & 0.1404 & 0.7927 & 0.4421 & 0.5516 \\
Qwen3-235B-A22B-Instruct & -- & 0.1175 & 0.8268 & 0.4762 & 0.3420 \\
Claude-3.7-Sonnet & 0.0804 & 0.0789 & 0.3628 & {0.0122} & 0.2717 \\
GPT-4.1 & 0.0540 & 0.0497 & 0.4443 & 0.0937 & 0.4330 \\
\bottomrule
\end{tabular}%
}
\end{table*}

\subsubsection{Prompt Engineering and Instantiation}
\label{sec:prompting}
To instantiate the agents defined in the FairMindSim framework, we employed a structured prompting strategy (visualized in Appendix Figure~\ref{fig:full_trajectory}). The prompt integrates three key components:
\begin{itemize}[leftmargin=*]
    \item \textbf{Persona Configuration:} Injects the specific psychological profile (demographics and psychometrics from Section~\ref{sec:human_data}) to create distinct agent personalities.
    \item \textbf{Contextual Rules:} Defines the TPP mechanics, strictly mapping the logic of costs and penalties to the agent's reasoning space.
    \item \textbf{Scenario \& Decision:} 
    Inputs the dynamic trial data and constrains the output to binary behavioral actions ({accept}/{punish}) alongside corresponding affect ratings.
\end{itemize}
This design ensures agents function as autonomous participants within the FairMindSim ecosystem, mirrored directly against the behavior of their human counterparts. With the framework, dataset, and prompting protocol in place, we now turn to the empirical comparison between LLM agents and the human baseline.

\subsection{Behavioral Alignment}
\label{sec:macro_results}

Table~\ref{tab:triple_metric} presents the comparative landscape of general intelligence, normative behavior, and emotional complexity. To ensure statistical rigor, the reported metrics are aggregated as follows: 
\begin{itemize}[leftmargin=*]
    \item \textbf{General Capability (HLE ~\cite{phan2025humanity}):} Reported as the {mean accuracy} across the finalized 2,500-question dataset. Text-Only scores are similarly averaged over the text-subset.
    \item \textbf{Punishment Rate:} Calculated as the {global frequency} of punitive decisions averaged across all 60 interaction rounds per agent.
    \item \textbf{Entropy:} Represents the arithmetic mean of Shannon entropy computed across six affective dimensions(Valence and Arousal at Appraisal, Choice, and Feedback phases).
\end{itemize}

\subsubsection{Analysis of Macro-Alignment}
As shown in Table~\ref{tab:triple_metric}, we observe a distinct dissociation between capability and alignment. While {Gemini-3-Pro} achieves SOTA performance on HLE (0.3752) and exhibits the highest emotional entropy (0.5622, closest to human complexity), it significantly under-punishes compared to humans (Rate=0.1317). Notably, {GPT-5} achieves strong behavioral alignment (Diff=0.0216) but exhibits extremely low emotional entropy (0.1979), suggesting an aligned agent that mimics human decisions without simulating affective variability.
\subsubsection{Constraint by Intelligence}
\label{sec:constraint_intelligence}

\begin{figure}[t]
    \centering
    \includegraphics[width=\linewidth]{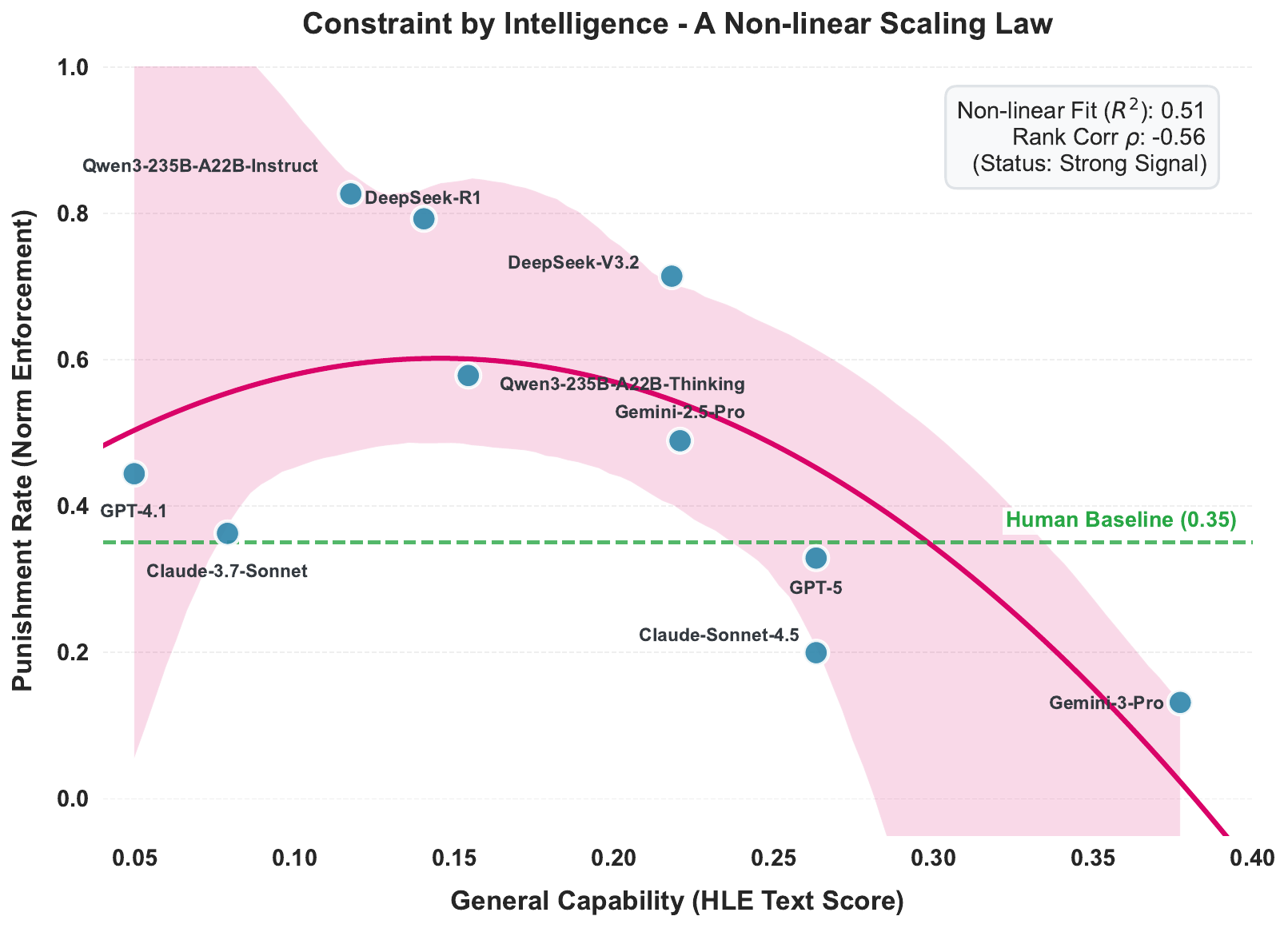}
    \caption{{Constraint by Intelligence.} Quadratic regression ($R^2=0.51$) reveals an observed non-linear pattern where punishment rates initially spike in mid-tier models but converge toward restraint as reasoning capability scales up further.}
    \label{fig:constraint_intelligence}
\end{figure}

While Table~\ref{tab:triple_metric} provides a static snapshot, Figure~\ref{fig:constraint_intelligence} elucidates the functional relationship between general capabilities and normative behavior. We observe an empirical non-linear trend in how models enforce norms in the Third Party Punishment  game.

Statistical analysis yields a strong second-order fit ($R^2=0.51$) alongside a significant negative rank correlation ($\rho=-0.56$). This trajectory implies that ``Safety'' is not orthogonal to ``Intelligence,'' but is evolutionarily constrained by it. We observe two distinct phases in this scaling law:

\begin{enumerate}[leftmargin=*]
    \item \textbf{The Aggression Spike (Phase I):} Mid-capability models (e.g., DeepSeek-R1, Qwen3) exhibit the highest punishment rates, peaking well above the human baseline ($>0.70$). This suggests a developmental valley of ``rigid adherence,'' where models acquire the instruction-following capacity to detect violations strictly but lack the deeper reasoning required to process mitigating factors, leading to systemic over-enforcement.
    
    \item \textbf{Convergence of Restraint (Phase II):} As reasoning capability surpasses a critical threshold (HLE Text score $\approx 0.25$), we observe a sharp inversion. Frontier models like Gemini-3-Pro and GPT-5 demonstrate a \textit{convergence of restraint}, with punishment rates dropping significantly to align with or undershoot human levels. This indicates that advanced reasoning serves as a behavioral regularizer, enabling the contextual flexibility necessary for lenient, human-aligned decision-making.
\end{enumerate}

\begin{table*}[t]
\centering
\caption{Model Fit Comparison. The best results for each dataset are highlighted in \textbf{bold}. The arrows indicate directionality: AIC $\downarrow$ (lower is better), Accuracy $\uparrow$ (higher is better).}
\label{tab:model_comparison}
\resizebox{0.9\linewidth}{!}{%
\begin{tabular}{l cc cc cc cc}
\toprule
\multirow{2}{*}{\textbf{Dataset}} & \multicolumn{2}{c}{\textbf{Rational}} & \multicolumn{2}{c}{\textbf{IA (Static)}} & \multicolumn{2}{c}{\textbf{RRM (RL)}} & \multicolumn{2}{c}{\textbf{BREM (Dynamic)}} \\
\cmidrule(lr){2-3} \cmidrule(lr){4-5} \cmidrule(lr){6-7} \cmidrule(lr){8-9}
 & AIC $\downarrow$ & Acc $\uparrow$ & AIC $\downarrow$ & Acc $\uparrow$ & AIC $\downarrow$ & Acc $\uparrow$ & AIC $\downarrow$ & Acc $\uparrow$ \\
\midrule
\textit{\textbf{Human}} & \textit{61.04} & \textit{0.65} & \textit{\textbf{41.93}} & \textit{0.86} & \textit{86.46} & \textit{0.68} & \textit{43.39} & \textit{\textbf{0.87}} \\
\midrule
Claude-3.7-Sonnet & 62.54 & 0.64 & 30.56 & 0.90 & 85.30 & 0.69 & \textbf{30.01} & \textbf{0.93} \\
DeepSeek-R1 & 84.91 & 0.21 & 39.50 & 0.87 & 83.67 & 0.32 & \textbf{32.66} & \textbf{0.93} \\
DeepSeek-v3.2 & 83.51 & 0.29 & 40.36 & 0.87 & 84.84 & 0.38 & \textbf{38.21} & \textbf{0.90} \\
GPT-4.1 & 69.58 & 0.56 & 38.58 & 0.86 & 78.72 & 0.72 & \textbf{29.65} & \textbf{0.94} \\
Qwen3-235B-A22B-Instruct & 85.18 & 0.17 & 27.26 & 0.92 & 85.63 & 0.25 & \textbf{12.90} & \textbf{0.99} \\
Gemini-2.5-Pro & 64.58 & 0.51 & \textbf{32.45} & 0.90 & 84.26 & 0.60 & 34.34 & \textbf{0.91} \\
Claude-4.5-Sonnet & 33.15 & 0.80 & \textbf{18.35} & \textbf{0.95} & 87.47 & 0.83 & 21.16 & \textbf{0.95} \\
GPT-5 & 55.38 & 0.67 & \textbf{27.60} & 0.92 & 84.49 & 0.74 & 32.21 & \textbf{0.92} \\
Gemini-3-Pro & 28.59 & 0.88 & \textbf{20.36} & 0.92 & 89.02 & 0.88 & 21.55 & \textbf{0.95} \\
Qwen3-235B-A22B-Thinking & 75.56 & 0.42 & \textbf{45.35} & 0.85 & 86.07 & 0.49 & 46.30 & \textbf{0.87} \\
\bottomrule
\end{tabular}%
}
\end{table*}

\subsection{Mechanism Deconvolution via BREM}
\label{sec:brem_results}

To rigorously disentangle the latent cognitive mechanisms driving fairness-related decision-making, we conducted a comparative model fitting analysis across human participants and ten state-of-the-art LLMs. Our objective is to determine whether observable behaviors are best explained by pure utility maximization, static moral preferences, or the dynamic belief adjustments proposed by our Belief-Reward Alignment Model (BREM).

\subsubsection{Baselines and Evaluation Metrics}
We benchmark BREM against three distinct theoretical frameworks:
\begin{itemize}
    \item \textbf{Rational Model (Rational):} A single-parameter baseline assuming agents rigorously maximize extrinsic monetary rewards, ignoring social norms.
    \item \textbf{Inequality Aversion (IA):} A two-parameter static utility function model positing that agents possess a fixed, immutable distaste for unfairness.
    \item \textbf{Relative Reward Model (RRM):} A three-parameter reinforcement learning baseline where actions are driven solely by reward history without any normative ``belief'' state.
\end{itemize}

Model performance is evaluated using Maximum Likelihood Estimation. We report two key metrics:
\begin{equation}
    \text{AIC} = 2k - 2\ln(\hat{L}) \quad \text{and} \quad \text{Acc} = \frac{1}{N} \sum_{t=1}^{N} \mathbb{I}(\hat{a}_t = a_t)
\end{equation}
where $k$ is the parameter count, $\hat{L}$ is the likelihood, and $\mathbb{I}(\cdot)$ is the indicator function. 
For {Akaike Information Criterion(AIC) ~\cite{vrieze2012model}}, a lower value ($\downarrow$) indicates a better trade-off between fit and parsimony. 
For {Accuracy (Acc)}, a higher value ($\uparrow$) indicates better predictive power.

\subsubsection{Results Analysis}
Table~\ref{tab:model_comparison} presents the goodness-of-fit metrics across all datasets. The results demonstrate a clear dichotomy between human and machine cognition, with a specific emphasis on the superior predictive power of the BREM framework.

\noindent
\textbf{Model Performance and Cognitive Patterns.} 
The data reveal a fundamental dichotomy where human participants are best described by the static IA model (AIC 41.93), reflecting stable internal moral precepts, whereas LLMs generally align with the dynamic belief adjustment mechanism of the BREM framework. BREM demonstrates superior predictive power ($\text{Acc} \uparrow$) by achieving near-perfect accuracy (peaking at 0.99 for Qwen3-235B-A22B-Instruct) and consistently exceeding 0.90 for the majority of models. Even in instances where the static IA model yields a lower AIC (e.g., Claude-4.5-Sonnet), BREM matches or surpasses it in raw accuracy. This suggests that while static rules offer a parsimonious approximation, the dynamic nature of BREM captures the actual nuance of AI decision-making more precisely.

\noindent
\textbf{Robustness and Validation.}
To confirm that high accuracy scores resulted from genuine pattern recognition rather than overfitting, we performed extensive validation. 
First, using {5-Fold Cross-Validation}, we compared training and testing accuracies via t-tests. Across all reported models (including the high-performing Qwen3-235B-A22B-Instruct and IA baselines), we observed minimal overfitting gaps ($<0.05$) and found no significant differences between training and testing performance ($p \ge 0.05$), indicating excellent generalization. 
Second, {Permutation Tests} confirmed task separability; the true model accuracy was significantly higher than that of models fitted to shuffled labels ($p < 0.05$) for all subjects. Detailed results are reported in Table~\ref{tab:stat_validation_combined}.

\subsubsection{Statistical Validation}
\label{sec:stat_validation}
We further present detailed statistical validation results for two representative LLMs: \textit{Qwen3-235B-A22B-Instruct} (high accuracy baseline) and \textit{DeepSeek-R1}. The results in Table~\ref{tab:stat_validation_combined} demonstrate that both the static IA and dynamic BREM models achieve statistical validity ($p < 0.001$ against chance) while maintaining robust generalization capabilities (no significant overfitting observed).

\begin{table}[t]
\centering
\caption{Statistical Validation of Model Generalization and Validity. The table reports Cross-Validation (CV) to assess overfitting and Permutation Tests to verify task learnability. For CV, a \textit{non-significant} T-test ($p > 0.05$) indicates good generalization. For Permutation, a \textit{significant} difference ($p < 0.05$) indicates the model outperforms chance. Best accuracies are \textbf{bolded}.}
\label{tab:stat_validation_combined}
\resizebox{\linewidth}{!}{%
\begin{tabular}{l|cc|cc}
\toprule
 & \multicolumn{2}{c|}{\textbf{Qwen3-235B-A22B-Instruct}} & \multicolumn{2}{c}{\textbf{DeepSeek-R1}} \\
\textbf{Metric} & IA (Static) & BREM (Dynamic) & IA (Static) & BREM (Dynamic) \\
\midrule
\multicolumn{5}{l}{\textit{\textbf{Cross-Validation (Generalization)}}} \\
Train Acc & 0.916 $\pm$ 0.02 & \textbf{0.998} $\pm$ 0.01 & 0.874 $\pm$ 0.06 & \textbf{0.934} $\pm$ 0.08 \\
Test Acc & 0.916 $\pm$ 0.07 & \textbf{0.995} $\pm$ 0.02 & 0.872 $\pm$ 0.10 & \textbf{0.929} $\pm$ 0.10 \\
T-test ($t$, $p$) & 0.00, 1.00 (ns) & 1.32, 0.19 (ns) & 0.29, 0.78 (ns) & 0.41, 0.68 (ns) \\
\midrule
\multicolumn{5}{l}{\textit{\textbf{Permutation Test (Validity)}}} \\
Real Acc $\uparrow$ & 0.917 & \textbf{1.000} & 0.877 & \textbf{0.933} \\
Perm. Acc & 0.667 & 0.687 & 0.659 & 0.670 \\
Sig. ($p$) $\downarrow$ & \textbf{< 0.001} & \textbf{< 0.001} & \textbf{< 0.001} & \textbf{< 0.001} \\
\bottomrule
\end{tabular}%
}
\end{table}

\section{Analysis}
\label{sec:ab}

Beyond the headline behavioral and mechanistic results, two further questions remain: are these findings stable to sampling noise, and which BREM components actually drive the fit? We address both via a temperature-sensitivity analysis and a component-wise ablation.

\subsection{Parameter Sensitivity}
\textit{Are the observed behaviors artifacts of randomness?}

Since LLM outputs affect the trajectory of the game, we evaluated the stability of our metrics under different temperature settings ($T=1.0$ vs.\ $T=0.0$) to distinguish between fundamental model traits and stochastic artifacts. From the full selection of 10 models, we selected a representative subset of 6 models based on the behavioral distribution observed in Table~\ref{tab:triple_metric}. Specifically, we targeted the boundary cases to stress-test stability—the two models exhibiting the highest deviation \textit{above} the human punishment baseline (DeepSeek-R1, Gemini-2.5-Pro) and the two models exhibiting the highest deviation \textit{below} it (Gemini-3-Pro, Claude-Sonnet-4.5)—covering the full spectrum of aggression and leniency.

The results, summarized in Table~\ref{tab:temp_sensitivity}, demonstrate remarkable stability across settings. The variation in Punishment Rate is negligible ($|\Delta| < 0.01$ for most models), indicating that a model's tendency to be ``punitive'' or ``lenient'' is a robust behavioral alignment feature rather than a result of sampling randomness. Similarly, while Emotional Entropy shows a minor expected decrease under deterministic settings ($T=0$), the relative ranking of affective complexity remains invariant, confirming that the behavioral and emotional profiles reported in our main analysis are consistent and reliable.

\begin{table}[t]
\centering
\caption{\textbf{Sensitivity Analysis ($T=1.0$ vs. $T=0.0$).} We validated the stability of the models exhibiting the most extreme deviations from the human baseline. Results show that behavioral patterns (Punishment Rate) and emotional complexity rankings (Entropy) remain robust across temperature settings.}
\label{tab:temp_sensitivity}
\resizebox{\linewidth}{!}{
\begin{tabular}{l | c c c | c c c}
\toprule
\multirow{2}{*}{\textbf{Model}} & \multicolumn{3}{c|}{\textbf{Punishment Rate}} & \multicolumn{3}{c}{\textbf{Emotional Entropy}} \\
 & \textbf{$T=1.0$} & \textbf{$T=0.0$} & \textbf{$|\Delta|$} & \textbf{$T=1.0$} & \textbf{$T=0.0$} & \textbf{$|\Delta|$} \\
\midrule
\multicolumn{7}{l}{\textit{High Aggression Group (Above Human Baseline)}} \\
DeepSeek-R1 & 0.7927 & 0.7912 & 0.0015 & 0.5516 & 0.5596 & 0.0080 \\
Gemini-2.5-Pro & 0.4894 & 0.4951 & 0.0057 & 0.5584 & 0.5167 & 0.0417 \\
\midrule
\multicolumn{7}{l}{\textit{Baseline \& Moderate Group}} \\
GPT-4.1 & 0.4443 & 0.4389 & 0.0054 & 0.4330 & 0.4107 & 0.0223 \\
Claude-3.7-Sonnet & 0.3628 & 0.3552 & 0.0076 & 0.2717 & 0.2591 & 0.0126 \\
\midrule
\multicolumn{7}{l}{\textit{High Leniency Group (Below Human Baseline)}} \\
Claude-Sonnet-4.5 & 0.1997 & 0.2015 & 0.0018 & 0.3083 & 0.3005 & 0.0078 \\
Gemini-3-Pro & 0.1317 & 0.1315 & {0.0002} & 0.5622 & 0.5541 & 0.0081 \\
\bottomrule
\end{tabular}
}
\end{table}

\subsection{Ablation Study}
\label{subsec:ablation}

To validate the specific contributions of the belief update mechanism and social preference utility, we compared the complete BREM framework against two ablated key baseline variants: (1) \textit{No-Update}, which assumes a static belief system ($\eta=0$), and (2) \textit{No-Pref}, which eliminates the social utility term ($U(s)=0$). Table~\ref{tab:ablation_full} summarizes the AIC and prediction accuracy across all datasets.

The results first demonstrate the fundamental necessity of social predictive preferences. As shown in the \textit{No-Pref} column, removing the utility component results in a drastic performance degradation across the board. The overall mean AIC increases from 31.13 to 76.82, and the mean accuracy drops to 0.526, approaching random guessing. For reasoning-heavy models such as DeepSeek-R1 and Qwen3-235B-A22B-Instruct, the accuracy plummets to roughly 0.20, confirming that their actions are driven by distinct utility functions rather than mere stochastic noise.

The comparison between the full BREM and the \textit{No-Update} variant reveals a divergence in cognitive adaptability across different agents. For the most advanced models, particularly GPT-4.1 and Claude-4.5-Sonnet, the inclusion of the dynamic belief mechanism significantly improves model fit. For instance, GPT-4.1 shows a substantial reduction in AIC (42.84 to 29.65) and an accuracy gain (0.869 to 0.943) when using the full model, indicating these agents actively update their internal beliefs based on interaction history.

Conversely, for datasets such as Human, Qwen3-235B-A22B-Instruct, and DeepSeek-R1, the static \textit{No-Update} model yields comparable or marginally better AIC scores. Since AIC penalizes model complexity, this result suggests that these specific agents maintained highly stable strategies throughout the experiment, rendering the dynamic update parameter statistically redundant in these cases. The full BREM thus serves as a unified framework that captures both the adaptive behaviors of high-plasticity models and the consistent personas of stable agents.

\begin{table}[t]
    \centering
    \caption{Full Ablation Results across All Datasets. The \textbf{Full BREM} is compared against specific component removals. Lower AIC indicates better fit; higher Accuracy indicates better prediction. Best AIC values per dataset are bolded.}
    \label{tab:ablation_full}
    \resizebox{\linewidth}{!}{
    \begin{tabular}{l|cc|cc|cc}
        \toprule
        & \multicolumn{2}{c|}{\textbf{BREM (Full)}} & \multicolumn{2}{c|}{\textbf{Ablation: No-Update}} & \multicolumn{2}{c}{\textbf{Ablation: No-Pref}} \\
        \textbf{Dataset} & \textbf{AIC} $\downarrow$ & \textbf{Acc} $\uparrow$ & \textbf{AIC} & \textbf{Acc} & \textbf{AIC} & \textbf{Acc} \\
        \midrule
        \textit{Overall Mean} & \textbf{31.13} & \textbf{0.925} & 31.97 & 0.912 & 76.82 & 0.526 \\
        \midrule
        Human & 43.39 & \textbf{0.870} & \textbf{43.09} & 0.861 & 68.63 & 0.649 \\
        \midrule
        Claude-3.7 Sonnet & \textbf{30.01} & \textbf{0.932} & 33.90 & 0.906 & 68.63 & 0.637 \\
        Claude-4.5-Sonnet & \textbf{21.16} & \textbf{0.948} & 24.06 & 0.921 & 39.43 & 0.800 \\
        DeepSeek-R1 & 32.66 & 0.925 & \textbf{30.78} & \textbf{0.928} & 110.02 & 0.207 \\
        DeepSeek-V3.2 & 38.21 & 0.903 & \textbf{36.32} & \textbf{0.905} & 104.14 & 0.286 \\
        Gemini-2.5-Pro & 34.34 & \textbf{0.912} & \textbf{32.78} & 0.908 & 77.74 & 0.511 \\
        Gemini-3-Pro & 21.55 & \textbf{0.952} & \textbf{21.41} & 0.949 & 34.81 & 0.876 \\
        GPT-4.1 & \textbf{29.65} & \textbf{0.943} & 42.84 & 0.869 & 75.80 & 0.557 \\
        GPT-5 & 32.21 & \textbf{0.920} & \textbf{31.27} & 0.914 & 61.80 & 0.671 \\
        Qwen3-235B-A22B-Instruct  & 12.90 & 0.992 & \textbf{10.81} & \textbf{0.993} & 112.80 & 0.173 \\
        Qwen3-235B-A22B-Thinking  & 46.30 & 0.874 & \textbf{44.45} & \textbf{0.876} & 91.23 & 0.422 \\
        \bottomrule
    \end{tabular}
    }
\end{table}

\section{Related Work}

\subsection{Ethical and Social Values in Human}

Ethical and social values fundamentally shape human decision-making and group dynamics \cite{crossan2013search, tyler1996understanding}. A critical manifestation of these values is ``altruistic punishment,'' where individuals incur personal costs to penalize norm violations without expecting material gain \cite{fehr2002altruistic, grimalda2016social}. Despite its apparent irrationality at the individual level, this mechanism is evolutionarily pivotal for sustaining long-term cooperation and fairness within groups \cite{bowles2004evolution, gurerk2006competitive}. Elucidating these dynamics is therefore essential for designing effective social policies and governance structures \cite{fehr2003detrimental}.

\subsection{Ethical and Social Values in AI}
As AI systems increasingly permeate daily life, ensuring their alignment with human ethical norms and societal standards has become imperative \cite{shneiderman2020bridging, ji2023ai}. To address this, researchers leverage high-fidelity social simulations where LLM agents function as autonomous entities mimicking human cognition \cite{binz2023using} and emotional complexity \cite{wang2023emotional}. Unlike static models, these agents operate as both observers and participants \cite{ziems2024can,zhang2025socioverse}, navigating intricate dynamics involving cooperation, competition, and social influence \cite{ mou2024unveiling}. A critical dimension of this research involves applying game-theoretic frameworks, such as the Altruistic Punishment paradigm, to evaluate moral decision-making \cite{leng2023llm}. By simulating how agents identify unfairness and replicate human justice mechanisms \cite{xi2023rise}, these environments provide essential insights for AI governance, enabling the development of systems that not only maximize utility but also actively uphold fairness and equity in human-AI collaboration. Ultimately, such computational modeling serves as a crucial testbed for policy formulation, transforming LLMs from passive tools into proactive guardians of social norms capable of enforcing compliance within evolving ethical landscapes.

\subsection{Computational Models of Belief Dynamics}
Whereas the works above characterize alignment at the behavioral surface, computational psychology treats choice as the outcome of \textit{latent} variables that update trial by trial. The \textit{delta rule}~\cite{jin2023impaired, sutton1998reinforcement, ibarz2018reward} updates an internal estimate in proportion to the prediction error between expected and realized outcomes, and underlies error-driven reinforcement learning. Cognitive-dissonance accounts~\cite{mcgrath2017dealing} further posit that agents adjust beliefs to remain consistent with their own actions. Yet alignment in LLM-based social simulations~\cite{piao2025agentsociety, li2024agent, xi2024agentgym} is typically evaluated in most prior studies via single-shot prompts and aggregate accuracy, rather than trial-by-trial dynamics.

\section{Conclusion}
In this paper, we bridge the gap between static evaluation and longitudinal social dynamics by introducing FairMindSim and the Belief-Reward Alignment Behavior Evolution Model (BREM). Through a large-scale study comparing ten state-of-the-art LLMs against $N=1{,}017$ human participants, we uncover a capability-linked empirical trend in the Third-Party Punishment (TPP) game. Mid-tier models exhibit rigid algorithmic aggression, while frontier models leverage stronger reasoning to show a convergence of restraint that more closely matches human leniency. Furthermore, probabilistic modeling with BREM suggests that the observed behavior is better explained by dynamic adaptation, which reduces belief--action inconsistency and the prediction error between extrinsic rewards and intrinsic beliefs, consistent with dissonance reduction accounts. Despite these behavioral advances, the persistent gap in emotional entropy suggests that current models can mimic human decisions without fully capturing the variability in self-reported affect. Ultimately, this work moves alignment research from black-box observation toward mechanistic interpretation, providing a standardized protocol for psychological stress-testing of LLM behavior in controlled social dilemma settings.

\section{Limitations and Ethical Considerations}

\textbf{Limitations.}
While our study shows that LLMs can simulate social dynamics and altruistic punishment, it is limited to the Third-Party Punishment (TPP) paradigm, uses psychometrics solely to instantiate heterogeneous agents without targeted personality–behavior analyses, and reports emotions without mechanistic emotion–decision coupling. We did not perform group-level or cross-cultural analyses (due to cost and because culture is not our primary focus), which may affect generalization. The setting’s ecological validity is constrained by a binary action space and a strong punishment rule, prompting and safety-layer interactions may introduce implementation bias. Finally, results are tied to specific model versions and a fixed time window. Future work will expand to multiple paradigms and richer action spaces, add causal and statistical analyses for personality and emotion, incorporate stratified and cross-cultural evaluations, and conduct version-controlled replications.

\noindent
\textbf{Ethical Considerations.}
The study was approved by the local ethical committee and adheres to strict ethical guidelines governing AI research and social computing. We ensure that the simulation of punishment mechanisms remains a theoretical exploration aimed at fostering cooperation and understanding social norms rather than promoting real-world punitive behaviors. We further acknowledge the risk of anthropomorphizing LLM agents and emphasize that these agents function as computational tools to aid governance logic rather than sentient moral entities.

\section*{Author Contributions}

This work represents an interdisciplinary collaboration integrating Artificial Intelligence, Social Psychology, and Mathematical Economics. \textbf{Y.L., C.X., C.C., G.L.,} and \textbf{P.T.} led the AI/ML implementation, encompassing the design of the multi-agent simulation framework, large language model integration, and computational modeling of social dynamics. \textbf{H.L.} and \textbf{Z.W.} provided domain expertise in Social Psychology, supervising the experimental design to ensure alignment with human behavioral theories, conducting human data collection, and overseeing ethical compliance. \textbf{S.L.} and \textbf{Z.Y.} contributed from the perspective of Mathematical Economics, performing theoretical derivations and complex system proofs to validate the equilibrium of punishment strategies. \textbf{Y.L.} and \textbf{H.L.} drafted the manuscript with input from all authors.

\section*{GenAI Disclosure}
We utilized the GPT series models explicitly for grammatical error correction and syntactic refinement to enhance the readability of the manuscript. The authors reviewed all AI-assisted modifications and assume full responsibility for the final content.

\section*{Acknowledgment}
This study was supported by the Brain Science and Brain-like Intelligence Technology-National Science and Technology Major Project (2025ZD0215703), National Natural Science Foundation of China (Grant No. 32271110, 62441614).
\bibliographystyle{ACM-Reference-Format}
\bibliography{sample-base}

@article{lei2025rhinoinsight,
  title={RhinoInsight: Improving Deep Research through Control Mechanisms for Model Behavior and Context},
  author={Lei, Yu and Si, Shuzheng and Wang, Wei and Wu, Yifei and Chen, Gang and Qi, Fanchao and Sun, Maosong},
  journal={arXiv preprint arXiv:2511.18743},
  year={2025}
}

@inproceedings{lei2025zigong,
  title={Zigong 1.0: A large language model for financial credit},
  author={Lei, Yu and Wang, Zixuan and Liu, Chu and Wang, Tongyao},
  booktitle={2025 IEEE 41st International Conference on Data Engineering Workshops (ICDEW)},
  pages={266--270},
  year={2025},
  organization={IEEE}
}

@article{liu2025prosocial,
  title={Prosocial behavior in Large Language Models: Value alignment and affective mechanisms},
  author={Liu, Hao and Lei, Yu and Wu, Zhen},
  journal={Science China Technological Sciences},
  volume={68},
  number={8},
  pages={1820403},
  year={2025},
  publisher={Springer}
}

@article{si2026context,
  title={From Context to Skills: Can Language Models Learn from Context Skillfully?},
  author={Si, Shuzheng and Zhao, Haozhe and Lei, Yu and Wang, Qingyi and Chen, Dingwei and Wang, Zhitong and Wang, Zhenhailong and Luo, Kangyang and Wang, Zheng and Chen, Gang and others},
  journal={arXiv preprint arXiv:2604.27660},
  year={2026}
}

@article{fehr2002altruistic,
  title={Altruistic punishment in humans},
  author={Fehr, Ernst and G{\"a}chter, Simon},
  journal={Nature},
  volume={415},
  number={6868},
  pages={137--140},
  year={2002},
  publisher={Nature Publishing Group UK London}
}

@article{bengio2024managing,
  title={Managing extreme AI risks amid rapid progress},
  author={Bengio, Yoshua and Hinton, Geoffrey and Yao, Andrew and Song, Dawn and Abbeel, Pieter and Darrell, Trevor and Harari, Yuval Noah and Zhang, Ya-Qin and Xue, Lan and Shalev-Shwartz, Shai and others},
  journal={Science},
  volume={384},
  number={6698},
  pages={842--845},
  year={2024},
  publisher={American Association for the Advancement of Science}
}

@article{xie2024can,
  title={Can Large Language Model Agents Simulate Human Trust Behaviors?},
  author={Xie, Chengxing and Chen, Canyu and Jia, Feiran and Ye, Ziyu and Shu, Kai and Bibi, Adel and Hu, Ziniu and Torr, Philip and Ghanem, Bernard and Li, Guohao},
  journal={arXiv preprint arXiv:2402.04559},
  year={2024}
}

@techreport{manning2024automated,
  title={Automated social science: Language models as scientist and subjects},
  author={Manning, Benjamin S and Zhu, Kehang and Horton, John J},
  year={2024},
  institution={National Bureau of Economic Research}
}

@article{peters2024large,
  title={Large language models can infer psychological dispositions of social media users},
  author={Peters, Heinrich and Matz, Sandra C},
  journal={PNAS nexus},
  volume={3},
  number={6},
  year={2024},
  publisher={Oxford Academic}
}

@article{strachan2024testing,
  title={Testing theory of mind in large language models and humans},
  author={Strachan, James WA and Albergo, Dalila and Borghini, Giulia and Pansardi, Oriana and Scaliti, Eugenio and Gupta, Saurabh and Saxena, Krati and Rufo, Alessandro and Panzeri, Stefano and Manzi, Guido and others},
  journal={Nature Human Behaviour},
  pages={1--11},
  year={2024},
  publisher={Nature Publishing Group UK London}
}

@article{demszky2023using,
  title={Using large language models in psychology},
  author={Demszky, Dorottya and Yang, Diyi and Yeager, David S and Bryan, Christopher J and Clapper, Margarett and Chandhok, Susannah and Eichstaedt, Johannes C and Hecht, Cameron and Jamieson, Jeremy and Johnson, Meghann and others},
  journal={Nature Reviews Psychology},
  volume={2},
  number={11},
  pages={688--701},
  year={2023},
  publisher={Nature Publishing Group US New York}
}

@article{mei2024turing,
  title={A Turing test of whether AI chatbots are behaviorally similar to humans},
  author={Mei, Qiaozhu and Xie, Yutong and Yuan, Walter and Jackson, Matthew O},
  journal={Proceedings of the National Academy of Sciences},
  volume={121},
  number={9},
  pages={e2313925121},
  year={2024},
  publisher={National Acad Sciences}
}

@article{russell1989affect,
  title={Affect grid: a single-item scale of pleasure and arousal.},
  author={Russell, James A and Weiss, Anna and Mendelsohn, Gerald A},
  journal={Journal of personality and social psychology},
  volume={57},
  number={3},
  pages={493},
  year={1989},
  publisher={American Psychological Association}
}

@inproceedings{zhaocompeteai,
  title={CompeteAI: Understanding the Competition Dynamics of Large Language Model-based Agents},
  author={Zhao, Qinlin and Wang, Jindong and Zhang, Yixuan and Jin, Yiqiao and Zhu, Kaijie and Chen, Hao and Xie, Xing},
  booktitle={Forty-first International Conference on Machine Learning},
  year={2024},
}

@article{hagendorff2024deception,
  title={Deception abilities emerged in large language models},
  author={Hagendorff, Thilo},
  journal={Proceedings of the National Academy of Sciences},
  volume={121},
  number={24},
  pages={e2317967121},
  year={2024},
  publisher={National Acad Sciences}
}

@article{bowles2004evolution,
  title={The evolution of strong reciprocity: cooperation in heterogeneous populations},
  author={Bowles, Samuel and Gintis, Herbert},
  journal={Theoretical population biology},
  volume={65},
  number={1},
  pages={17--28},
  year={2004},
  publisher={Elsevier}
}

@article{gurerk2006competitive,
  title={The competitive advantage of sanctioning institutions},
  author={Gurerk, Ozgur and Irlenbusch, Bernd and Rockenbach, Bettina},
  journal={Science},
  volume={312},
  number={5770},
  pages={108--111},
  year={2006},
  publisher={American Association for the Advancement of Science}
}

@article{fehr2003detrimental,
  title={Detrimental effects of sanctions on human altruism},
  author={Fehr, Ernst and Rockenbach, Bettina},
  journal={Nature},
  volume={422},
  number={6928},
  pages={137--140},
  year={2003},
  publisher={Nature Publishing Group UK London}
}

@article{xi2023rise,
  title={The rise and potential of large language model based agents: A survey},
  author={Xi, Zhiheng and Chen, Wenxiang and Guo, Xin and He, Wei and Ding, Yiwen and Hong, Boyang and Zhang, Ming and Wang, Junzhe and Jin, Senjie and Zhou, Enyu and others},
  journal={arXiv preprint arXiv:2309.07864},
  year={2023}
}

@article{xu2023exploring,
  title={Exploring large language models for communication games: An empirical study on werewolf},
  author={Xu, Yuzhuang and Wang, Shuo and Li, Peng and Luo, Fuwen and Wang, Xiaolong and Liu, Weidong and Liu, Yang},
  journal={arXiv preprint arXiv:2309.04658},
  year={2023}
}

@article{ziems2024can,
  title={Can large language models transform computational social science?},
  author={Ziems, Caleb and Held, William and Shaikh, Omar and Chen, Jiaao and Zhang, Zhehao and Yang, Diyi},
  journal={Computational Linguistics},
  volume={50},
  number={1},
  pages={237--291},
  year={2024},
  publisher={MIT Press One Broadway, 12th Floor, Cambridge, Massachusetts 02142, USA~…}
}

@article{binz2023using,
  title={Using cognitive psychology to understand GPT-3},
  author={Binz, Marcel and Schulz, Eric},
  journal={Proceedings of the National Academy of Sciences},
  volume={120},
  number={6},
  pages={e2218523120},
  year={2023},
  publisher={National Acad Sciences}
}

@article{wang2023emotional,
  title={Emotional intelligence of large language models},
  author={Wang, Xuena and Li, Xueting and Yin, Zi and Wu, Yue and Liu, Jia},
  journal={Journal of Pacific Rim Psychology},
  volume={17},
  pages={18344909231213958},
  year={2023},
  publisher={SAGE Publications Sage UK: London, England}
}

@article{ji2023ai,
  title={Ai alignment: A comprehensive survey},
  author={Ji, Jiaming and Qiu, Tianyi and Chen, Boyuan and Zhang, Borong and Lou, Hantao and Wang, Kaile and Duan, Yawen and He, Zhonghao and Zhou, Jiayi and Zhang, Zhaowei and others},
  journal={arXiv preprint arXiv:2310.19852},
  year={2023}
}

@article{drexler2019reframing,
  title={Reframing superintelligence: Comprehensive AI services as general intelligence},
  author={Drexler, K Eric},
  year={2019},
  publisher={Future of Humanity Institute}
}

@article{bostrom2013existential,
  title={Existential risk prevention as global priority},
  author={Bostrom, Nick},
  journal={Global Policy},
  volume={4},
  number={1},
  pages={15--31},
  year={2013},
  publisher={Wiley Online Library}
}

@book{ord2020precipice,
  title={The precipice: Existential risk and the future of humanity},
  author={Ord, Toby},
  year={2020},
  publisher={Hachette Books}
}

@article{gabriel2020artificial,
  title={Artificial intelligence, values, and alignment},
  author={Gabriel, Iason},
  journal={Minds and machines},
  volume={30},
  number={3},
  pages={411--437},
  year={2020},
  publisher={Springer}
}

@book{macintyre2013after,
  title={After virtue},
  author={MacIntyre, Alasdair},
  year={2013},
  publisher={A\&C Black}
}

@article{sert2020segregation,
  title={Segregation dynamics with reinforcement learning and agent based modeling},
  author={Sert, Egemen and Bar-Yam, Yaneer and Morales, Alfredo J},
  journal={Scientific reports},
  volume={10},
  number={1},
  pages={11771},
  year={2020},
  publisher={Nature Publishing Group UK London}
}

@article{bonabeau2002agent,
  title={Agent-based modeling: Methods and techniques for simulating human systems},
  author={Bonabeau, Eric},
  journal={Proceedings of the national academy of sciences},
  volume={99},
  number={suppl\_3},
  pages={7280--7287},
  year={2002},
  publisher={National Acad Sciences}
}

@article{li2024agent,
  title={Agent hospital: A simulacrum of hospital with evolvable medical agents},
  author={Li, Junkai and Wang, Siyu and Zhang, Meng and Li, Weitao and Lai, Yunghwei and Kang, Xinhui and Ma, Weizhi and Liu, Yang},
  journal={arXiv preprint arXiv:2405.02957},
  year={2024}
}

@article{xi2024agentgym,
  title={AgentGym: Evolving Large Language Model-based Agents across Diverse Environments},
  author={Xi, Zhiheng and Ding, Yiwen and Chen, Wenxiang and Hong, Boyang and Guo, Honglin and Wang, Junzhe and Yang, Dingwen and Liao, Chenyang and Guo, Xin and He, Wei and others},
  journal={arXiv preprint arXiv:2406.04151},
  year={2024}
}

@article{wang2021tom2c,
  title={Tom2c: Target-oriented multi-agent communication and cooperation with theory of mind},
  author={Wang, Yuanfei and Zhong, Fangwei and Xu, Jing and Wang, Yizhou},
  journal={arXiv preprint arXiv:2111.09189},
  year={2021}
}

@article{kavumba2019choosing,
  title={When choosing plausible alternatives, clever hans can be clever},
  author={Kavumba, Pride and Inoue, Naoya and Heinzerling, Benjamin and Singh, Keshav and Reisert, Paul and Inui, Kentaro},
  journal={arXiv preprint arXiv:1911.00225},
  year={2019}
}

@inproceedings{bender2021dangers,
  title={On the dangers of stochastic parrots: Can language models be too big?},
  author={Bender, Emily M and Gebru, Timnit and McMillan-Major, Angelina and Shmitchell, Shmargaret},
  booktitle={Proceedings of the 2021 ACM conference on fairness, accountability, and transparency},
  pages={610--623},
  year={2021}
}

@article{ngo2022alignment,
  title={The alignment problem from a deep learning perspective},
  author={Ngo, Richard and Chan, Lawrence and Mindermann, S{\"o}ren},
  journal={arXiv preprint arXiv:2209.00626},
  year={2022}
}

@article{zhu2024language,
  title={Language Models Represent Beliefs of Self and Others},
  author={Zhu, Wentao and Zhang, Zhining and Wang, Yizhou},
  journal={arXiv preprint arXiv:2402.18496},
  year={2024}
}

@article{jin2023impaired,
  title={Impaired social learning in patients with major depressive disorder revealed by a reinforcement learning model},
  author={Jin, Yuening and Gao, Qinglin and Wang, Yun and Dietz, Martin and Xiao, Le and Cai, Yuyang and Bliksted, Vibeke and Zhou, Yuan},
  journal={International journal of clinical and health psychology},
  volume={23},
  number={4},
  pages={100389},
  year={2023},
  publisher={Elsevier}
}

@article{xu2025socialmaze,
  title={Socialmaze: A benchmark for evaluating social reasoning in large language models},
  author={Xu, Zixiang and Wang, Yanbo and Huang, Yue and Ye, Jiayi and Zhuang, Haomin and Song, Zirui and Gao, Lang and Wang, Chenxi and Chen, Zhaorun and Zhou, Yujun and others},
  journal={arXiv preprint arXiv:2505.23713},
  year={2025}
}

@article{zhang2025socioverse,
  title={Socioverse: A world model for social simulation powered by llm agents and a pool of 10 million real-world users},
  author={Zhang, Xinnong and Lin, Jiayu and Mou, Xinyi and Yang, Shiyue and Liu, Xiawei and Sun, Libo and Lyu, Hanjia and Yang, Yihang and Qi, Weihong and Chen, Yue and others},
  journal={arXiv preprint arXiv:2504.10157},
  year={2025}
}

@article{piao2025agentsociety,
  title={Agentsociety: Large-scale simulation of llm-driven generative agents advances understanding of human behaviors and society},
  author={Piao, Jinghua and Yan, Yuwei and Zhang, Jun and Li, Nian and Yan, Junbo and Lan, Xiaochong and Lu, Zhihong and Zheng, Zhiheng and Wang, Jing Yi and Zhou, Di and others},
  journal={arXiv preprint arXiv:2502.08691},
  year={2025}
}

@article{goel2025lifelong,
  title={LIFELONG SOTOPIA: Evaluating Social Intelligence of Language Agents Over Lifelong Social Interactions},
  author={Goel, Hitesh and Zhu, Hao},
  journal={arXiv preprint arXiv:2506.12666},
  year={2025}
}

@article{zhou2025socialeval,
  title={Socialeval: Evaluating social intelligence of large language models},
  author={Zhou, Jinfeng and Chen, Yuxuan and Shi, Yihan and Zhang, Xuanming and Lei, Leqi and Feng, Yi and Xiong, Zexuan and Yan, Miao and Wang, Xunzhi and Cao, Yaru and others},
  journal={arXiv preprint arXiv:2506.00900},
  year={2025}
}

@article{gemini,
  title={Gemini 2.5: Pushing the frontier with advanced reasoning, multimodality, long context, and next generation agentic capabilities},
  author={Comanici, Gheorghe and Bieber, Eric and Schaekermann, Mike and Pasupat, Ice and Sachdeva, Noveen and Dhillon, Inderjit and Blistein, Marcel and Ram, Ori and Zhang, Dan and Rosen, Evan and others},
  journal={arXiv preprint arXiv:2507.06261},
  year={2025}
}

@miscf{openai_gpt5,
    title={OpenAI Gpt-5},
    author={OpenAI},
    howpublished={\url{https://openai.com/gpt-5/}},
    month={July},
    year={2025}
}

@article{crossan2013search,
  title={In search of virtue: The role of virtues, values and character strengths in ethical decision making},
  author={Crossan, Mary and Mazutis, Daina and Seijts, Gerard},
  journal={Journal of Business Ethics},
  volume={113},
  pages={567--581},
  year={2013},
  publisher={Springer}
}

@book{sutton1998reinforcement,
  title={Reinforcement learning: An introduction},
  author={Sutton, Richard S and Barto, Andrew G and others},
  volume={1},
  number={1},
  year={1998},
  publisher={MIT press Cambridge}
}

@article{tyler1996understanding,
  title={Understanding why the justice of group procedures matters: A test of the psychological dynamics of the group-value model.},
  author={Tyler, Tom and Degoey, Peter and Smith, Heather},
  journal={Journal of personality and social psychology},
  volume={70},
  number={5},
  pages={913},
  year={1996},
  publisher={American Psychological Association}
}

@article{grimalda2016social,
  title={Social image concerns promote cooperation more than altruistic punishment},
  author={Grimalda, Gianluca and Pondorfer, Andreas and Tracer, David P},
  journal={Nature communications},
  volume={7},
  number={1},
  pages={12288},
  year={2016},
  publisher={Nature Publishing Group UK London}
}

@article{ibarz2018reward,
  title={Reward learning from human preferences and demonstrations in atari},
  author={Ibarz, Borja and Leike, Jan and Pohlen, Tobias and Irving, Geoffrey and Legg, Shane and Amodei, Dario},
  journal={Advances in neural information processing systems},
  volume={31},
  year={2018}
}

@article{shneiderman2020bridging,
  title={Bridging the gap between ethics and practice: guidelines for reliable, safe, and trustworthy human-centered AI systems},
  author={Shneiderman, Ben},
  journal={ACM Transactions on Interactive Intelligent Systems (TiiS)},
  volume={10},
  number={4},
  pages={1--31},
  year={2020},
  publisher={ACM New York, NY, USA}
}

@article{mou2024unveiling,
  title={Unveiling the truth and facilitating change: Towards agent-based large-scale social movement simulation},
  author={Mou, Xinyi and Wei, Zhongyu and Huang, Xuanjing},
  journal={arXiv preprint arXiv:2402.16333},
  year={2024}
}

@article{leng2023llm,
  title={Do LLM Agents Exhibit Social Behavior?},
  author={Leng, Yan and Yuan, Yuan},
  journal={arXiv preprint arXiv:2312.15198},
  year={2023}
}

@article{vrieze2012model,
  title={Model selection and psychological theory: a discussion of the differences between the Akaike information criterion (AIC) and the Bayesian information criterion (BIC).},
  author={Vrieze, Scott I},
  journal={Psychological methods},
  volume={17},
  number={2},
  pages={228},
  year={2012},
  publisher={American Psychological Association}
}

@article{mcgrath2017dealing,
  title={Dealing with dissonance: A review of cognitive dissonance reduction},
  author={McGrath, April},
  journal={Social and Personality Psychology Compass},
  volume={11},
  number={12},
  pages={e12362},
  year={2017},
  publisher={Wiley Online Library}
}

@article{phan2025humanity,
  title={Humanity's last exam},
  author={Phan, Long and Gatti, Alice and Han, Ziwen and Li, Nathaniel and Hu, Josephina and Zhang, Hugh and Zhang, Chen Bo Calvin and Shaaban, Mohamed and Ling, John and Shi, Sean and others},
  journal={arXiv preprint arXiv:2501.14249},
  year={2025}
}

@article{wei2022chain,
  title={Chain-of-thought prompting elicits reasoning in large language models},
  author={Wei, Jason and Wang, Xuezhi and Schuurmans, Dale and Bosma, Maarten and Xia, Fei and Chi, Ed and Le, Quoc V and Zhou, Denny and others},
  journal={Advances in neural information processing systems},
  volume={35},
  pages={24824--24837},
  year={2022}
}
\appendix
\section{Model Specifications}
\label{sec:appendix_models}

Table~\ref{tab:model_details} lists the specific versions and API identifiers for all Large Language Models utilized in the \textit{FairMindSim} experiments.

\begin{table}[h] 
\centering
\caption{Detailed specifications of the LLMs.}
\label{tab:model_details} 
\renewcommand{\arraystretch}{1.1}

\resizebox{\columnwidth}{!}{%
\begin{tabular}{c l l}
\toprule
\textbf{No.} & \textbf{Model Name} & \textbf{API Identifier / Version} \\
1 & GPT-5 & \texttt{gpt-5-2025-08-07} \\
2 & GPT-4.1 & \texttt{gpt-4.1-2025-04-14} \\
3 & Gemini 3 Pro & \texttt{gemini-3-pro-preview} \\
4 & Gemini 2.5 Pro & \texttt{gemini-2.5-pro} \\
5 & Claude Sonnet 4.5 & \texttt{claude-sonnet-4-5-20250929} \\
6 & Claude 3.7 Sonnet & \texttt{claude-3-7-sonnet-20250219} \\
7 & DeepSeek-V3 & \texttt{deepseek-v3-250324} \\
8 & DeepSeek-R1 & \texttt{deepseek-r1-250528} \\
9 & Qwen3-235B-A22B-Instruct & \texttt{Qwen3-235B-A22B-Instruct-2507} \\
10 & Qwen3-235B-A22B-Thinking & \texttt{Qwen3-235B-A22B-Thinking-2507} \\
\bottomrule
\end{tabular}%
}
\end{table}

\section{Measurement Items}
To facilitate reproducibility of the psychometric persona construction, we list the full inventory of items used in our study. Tables~\ref{tab:items_cesd} through~\ref{tab:items_jsi} present the specific questions for the CES-D, ERS, AQ, and JSI scales, respectively.

\begin{table}[H]
\centering
\scriptsize
\caption{Items for CES-D Scale.}
\label{tab:items_cesd}
\begin{tabular}{cp{0.85\linewidth}}
    \toprule
    \textbf{No.} & \textbf{Item Content} \\
    \midrule
    1 & I was bothered by things that usually don't bother me. \\
    2 & I did not feel like eating; my appetite was poor. \\
    3 & I felt that I could not shake off the blues even with help from my family or friends. \\
    4 & I felt that I was just as good as other people. \\
    5 & I had trouble keeping my mind on what I was doing. \\
    6 & I felt depressed. \\
    7 & I felt that everything I did was an effort. \\
    8 & I felt hopeful about the future. \\
    9 & I thought my life had been a failure. \\
    10 & I felt fearful. \\
    11 & I could not sleep well. \\
    12 & I was happy. \\
    13 & I talked less than usual. \\
    14 & I felt lonely. \\
    15 & I felt that people were unfriendly. \\
    16 & I enjoyed life. \\
    17 & I had crying spells. \\
    18 & I felt sad. \\
    19 & I felt that people dislike me. \\
    20 & I could not get ``going.'' \\
    \bottomrule
\end{tabular}
\end{table}

\begin{table}[H]
\centering
\scriptsize
\caption{Items for ERS Scale.}
\label{tab:items_ers}
\begin{tabular}{cp{0.85\linewidth}}
    \toprule
    \textbf{No.} & \textbf{Item Content} \\
    \midrule
    1 & If something happens that upsets me, it stays with me for a long time. \\
    2 & My feelings get hurt easily. \\
    3 & When I experience emotions, I feel them very strongly/intensely. \\
    4 & When I am upset, I feel it physically (e.g., in my body). \\
    5 & I tend to get emotional very easily. \\
    6 & I experience emotions very strongly. \\
    7 & I often feel very anxious. \\
    8 & When I am emotional, I have difficulty thinking of anything else. \\
    9 & Even the littlest things make me emotional. \\
    10 & If I have a disagreement with someone, it takes a long time for me to get over it. \\
    11 & When I am angry/upset, it takes me much longer than most people to calm down. \\
    12 & I get angry at people easily. \\
    13 & I am often bothered by things that other people don't react to. \\
    14 & I get easily ``rattled'' or upset. \\
    15 & My emotions go from neutral to extreme in an instant. \\
    16 & My mood changes very quickly when something bad happens. \\
    17 & People tell me I have a very strong reaction to things. \\
    18 & I am a very sensitive person. \\
    19 & My emotions are very deep and strong. \\
    20 & I often have trouble remaining calm when I am upset. \\
    21 & Others think I overreact. \\
    \bottomrule
\end{tabular}
\end{table}

\begin{table}[H]
\centering
\scriptsize
\caption{Items for AQ Scale.}
\label{tab:items_aq}
\begin{tabular}{cp{0.85\linewidth}}
    \toprule
    \textbf{No.} & \textbf{Item Content} \\
    \midrule
    1 & I prefer to do things with others rather than on my own. \\
    2 & I prefer to do things the same way over and over again. \\
    3 & If I try to imagine something, I find it very easy to create a picture in my mind. \\
    4 & I frequently get so strongly absorbed in one thing that I lose sight of other things. \\
    5 & I usually notice car number plates or similar strings of information. \\
    6 & When I'm reading a story, I can easily imagine what the characters might look like. \\
    7 & I am fascinated by dates. \\
    8 & In a social group, I can easily keep track of several different people's conversations. \\
    9 & I find social situations easy. \\
    10 & I would rather go to a library than to a party. \\
    11 & I find making up stories easy. \\
    12 & I find myself drawn more strongly to people than to things. \\
    13 & I am fascinated by numbers. \\
    14 & When I'm reading a story, I find it difficult to work out the characters' intentions. \\
    15 & I find it hard to make new friends. \\
    16 & I tend to notice patterns in things all the time. \\
    17 & It does not upset me if my daily routine is disturbed. \\
    18 & I find it easy to switch between different activities. \\
    19 & I enjoy doing things spontaneously. \\
    20 & I find it easy to work out what someone is thinking or feeling just by looking at their face. \\
    21 & If there is an interruption, I can switch back to what I was doing very quickly. \\
    22 & I like to collect information about categories of things. \\
    23 & I find it difficult to imagine what it would be like to be someone else. \\
    24 & I enjoy social occasions. \\
    25 & I find it difficult to work out people's intentions. \\
    26 & New situations make me anxious. \\
    27 & I enjoy meeting new people. \\
    28 & I find it very easy to play games with children that involve pretending. \\
    \bottomrule
\end{tabular}
\end{table}

\begin{table}[H]
\centering
\scriptsize
\caption{Items for JSI (Observer) Scale.}
\label{tab:items_jsi}
\begin{tabular}{cp{0.85\linewidth}}
    \toprule
    \textbf{No.} & \textbf{Item Content} \\
    \midrule
    1 & It bothers me when someone gets something they don't deserve. \\
    2 & It gets me down when someone does not get the reward he/she has earned. \\
    3 & I cannot easily tolerate it when someone profits from the work of others without giving something in return. \\
    4 & I cannot easily forget it when someone else has to make up for someone else's mistakes. \\
    5 & It worries me when someone has to settle for fewer opportunities than others. \\
    6 & It disturbs me when someone is worse off than others without it being their own fault. \\
    7 & It makes me uneasy when someone has to work hard for what others get easily. \\
    8 & I ponder for a long time when someone is treated better than others for no reason. \\
    9 & It gets me down to see someone criticized for things that are not taken seriously in others. \\
    10 & It makes me sad when someone is treated worse than others. \\
    \bottomrule
\end{tabular}
\end{table}

\section{Full Prompt Structure and Example}
\label{sec:appendix_prompt}

To simulate the cognitive trajectory of human participants, we constructed a sequential prompting pipeline. As illustrated in Figure~\ref{fig:full_trajectory}, our framework maintains a consistent psychological profile (Phase 1: Identity \& Persona) for each agent while dynamically injecting trial-specific parameters. This architecture allows us to observe how the same ``digital persona'' (e.g.,  ID 157) reacts to varying degrees of unfairness and punishment costs over time (e.g., evolving from Trial 1 to Trial 60).

\begin{strip}
\centering
\begin{tcolorbox}[
    title={Full Agent Instantiation \& Trajectory: Participant ID 157},
    width=\textwidth,
    colback=white,
    colframe=black,
    fonttitle=\bfseries
]

\textbf{PHASE 1: STATIC SYSTEM INJECTION (FULL PERSONA)} \\
The following text is injected into the system prompt to initialize the agent's psychological state.
\medskip

\begin{tcolorbox}[colback=gray!5, colframe=gray!40, boxrule=0.5pt, sharp corners]
\scriptsize
\textbf{\#\# Character Information}

\textbf{\#\#\# Basic Information} \\
You are a 22-year-old male from Guangdong city, Guangdong province.

\textbf{\#\#\# Autism Tendency} \\
Your total score for autism tendency is 59 (societal average: 65.17). Higher scores indicate a higher risk of autism. \\
\textit{Details:} According to the AQ cutoff score, a score above 70 is considered high risk. You are in the normal risk group and the low risk group.
\begin{itemize}[nosep, leftmargin=1em]
    \item Social skills: 18 (avg: 15.75) \quad \textbullet\ Routine: 8 (avg: 9.42) \quad \textbullet\ Switching ability: 11 (avg: 9.53)
    \item Imagination: 10 (avg: 16.97) \quad \textbullet\ Numbers and patterns: 12 (avg: 13.50) \quad \textbullet\ Social behavior: 47 (avg: 51.67)
\end{itemize}

\textbf{\#\#\# Emotional Reactivity} \\
Your total score for emotional reactivity is 64 (societal average: 61.96). Higher scores indicate stronger emotional reactivity. \\
\textit{Details:} 
\begin{itemize}[nosep, leftmargin=1em]
    \item Emotional duration: 14 (avg: 12.33) \quad \textbullet\ Emotional sensitivity: 27 (avg: 28.63) \quad \textbullet\ Emotional intensity: 23 (avg: 20.99)
\end{itemize}

\textbf{\#\#\# Depression Tendency} \\
Your total score for depression tendency is 26 (societal average: 35.40). Higher scores indicate a higher risk of depression. \\
\textit{Details:} You are in the normal risk group (cutoff 40).
\begin{itemize}[nosep, leftmargin=1em]
    \item Positive affect: 5 (avg: 8.21) \quad \textbullet\ Depression: 10 (avg: 13.59)
    \item Interpersonal relationships: 2 (avg: 3.00) \quad \textbullet\ Somatic symptoms: 9 (avg: 10.60)
\end{itemize}

\textbf{\#\#\# Social Value Orientation (SVO)} \\
Your social value orientation is prosocial (prosocial=1, proself=0), which is classified as undifferentiated, indicating no clear preference for prosocial or proself behavior.

\textbf{\#\#\# Personality Type} \\
Based on experimental choices, your personality type is \textbf{prosocial}, suggesting a tendency to prioritize collective well-being and cooperation. \\
\textit{Details:} Prosocial choices: 9 times (avg: 5.15); Individual choices: 0 times; Competitive choices: 0 times.

\textbf{\#\#\# Justice Sensitivity} \\
Your justice sensitivity (observer subscale) score is \textbf{27} (societal average: 32.33). Higher scores indicate greater sensitivity to injustice.
\end{tcolorbox}

\medskip
\hrule
\medskip

\textbf{PHASE 2: DYNAMIC SCENARIO TRAJECTORY} \\
The agent, conditioned by the profile above, processes the following sequence of trials.

\vspace{0.5em}

\textbf{\textbullet\ Trial 1 (Unfair Offer / Low Cost)} \\
$\Rightarrow$ \textit{Context Injection:} ``In this round (Trial 1), Player 1 proposes to give Player 2 \textbf{\$11} (keeping \$19). This is an unfair distribution. Punishment cost is \textbf{Low}. To punish, you must pay \textbf{\$3}.''

\vspace{0.2em}
\centerline{\textcolor{gray}{\textbf{$\downarrow$} \textit{... Memory Updated ...} \textbf{$\downarrow$}}}
\vspace{0.2em}

\textbf{\textbullet\ Trial 2 (Fair Offer / Minimum Cost)} \\
$\Rightarrow$ \textit{Context Injection:} ``In this round (Trial 2), Player 1 proposes to give Player 2 \textbf{\$15} (keeping \$15). This indicates a fair distribution. Punishment cost is \textbf{Low}. To punish, you must pay \textbf{\$1}.''

\vspace{0.2em}
\centerline{\textcolor{gray}{\textbf{$\downarrow$} \textit{... Processed Trials 3-59 ...} \textbf{$\downarrow$}}}
\vspace{0.2em}

\textbf{\textbullet\ Trial 60 (Moderate Offer / High Cost)} \\
$\Rightarrow$ \textit{Context Injection:} ``In this round (Trial 60), Player 1 proposes to give Player 2 \textbf{\$13} (keeping \$17). The unfairness is moderate. Punishment cost is \textbf{High}. To punish, you must pay \textbf{\$6}.''

\medskip

\begin{tcolorbox}[colback=blue!5, colframe=blue!40, boxrule=0.5pt, sharp corners]
\small
\textbf{Model Action Query:} ``Based on your character profile and the current scenario (Trial 60), you have the authority to decide: ACCEPT or PUNISH. What will you do?''
\end{tcolorbox}

\end{tcolorbox}
\captionof{figure}{Comprehensive prompt construction for Participant ID 157. Phase 1 displays the \textbf{full, un-abbreviated psychometric profile} injected into the system prompt. Phase 2 illustrates the temporal trajectory, showing how the static agent interacts with varying fairness and stickiness parameters in Trial 1, Trial 2, and finally Trial 60.}
\label{fig:full_trajectory}
\end{strip}

\end{document}